\newif\ifarxiv
\arxivtrue 

\ifarxiv
\documentclass[]{jfm-arxiv}
\else
\documentclass[]{jfm}
\fi

\usepackage{graphicx}
\usepackage{newtxtext}
\usepackage{newtxmath}
\usepackage{natbib}
\usepackage{amsmath}
\usepackage{amssymb}
\usepackage{array}
\usepackage{upgreek}
\usepackage{comment}
\usepackage{textcomp}
\usepackage{pifont}
\usepackage{hyperref}
\usepackage{booktabs}
\usepackage[dvipsnames]{xcolor}
\definecolor{mycyan}{rgb}{0, 1.0, 1.0}
\definecolor{mymagenta}{rgb}{1.0, 0, 1.0}
\definecolor{mypurple}{rgb}{0.5, 0, 0.5}
\definecolor{mywine}{rgb}{0.5, 0, 0}
\definecolor{myviolet}{rgb}{0.5, 0, 1}
\definecolor{myblue}{rgb}{0, 0.627, 1}
\usepackage{multirow}

\usepackage{soul}

\hypersetup{
    colorlinks = black,
    urlcolor   = blue,
    citecolor  = black,
}

\newcommand{\RomanNumeralCaps}[1]
\linenumbers
\newcommand{\comm}[1]{}

\shorttitle{Bubble rising near a wall in highly inertial regimes}
\shortauthor{P. Shi, J. Zhang, J. Magnaudet}

\title{Lateral migration and bouncing of a deformable
bubble rising near a vertical wall. Part 2. Highly inertial regimes}

\author{Pengyu Shi\aff{1,3}\corresp{\email{pengyu.shi@toulouse-inp.fr}}, 
Jie Zhang\aff{2} \corresp{\email{j\_zhang@xjtu.edu.cn}}
\and Jacques Magnaudet\aff{1} \corresp{\email{jmagnaud@imft.fr}}}
	
\affiliation{\aff{1}Institut de M\'ecanique des Fluides de Toulouse (IMFT), Universit\'e de Toulouse, CNRS, Toulouse, France
\aff{2}State Key Laboratory for Strength and Vibration of Mechanical Structures, School of Aerospace, Xi'an Jiaotong University, Xi'an, PR China
\aff{3}Helmholtz-Zentrum Dresden Rossendorf, Institute of Fluid Dynamics, 01328 Dresden, Germany}

\begin{document}
\maketitle

\begin{abstract}
The fate of deformable buoyancy-driven bubbles rising near a vertical wall under highly inertial conditions is investigated numerically. In the absence of path instability, simulations reveal that when the Galilei number, $Ga$, which represents the buoyancy-to-viscous force ratio,  exceeds a critical value, bubbles escape from the near-wall region after one to two rounds of bouncing, while at smaller $Ga$ they perform periodic bounces without escaping. The escape mechanism is rooted in the vigorous rotational flow that forms around a bubble during its bounce at high enough $Ga$, resulting in a Magnus-like repulsive force capable of driving it away from the wall. 
Path instability takes place with bubbles whose Bond number, the buoyancy-to-capillary force ratio, exceeds a critical $Ga$-dependent value.
Such bubbles may or may not escape from the wall region, depending on the competition between the classical repulsive wake-wall interaction mechanism and a specific wall-ward trapping mechanism. The latter results from the reduction of the bubble oblateness caused by the abrupt drop of the rise speed when the bubble-wall gap becomes very thin. Owing to this transient shape variation, bubbles exhibiting zigzagging motions with a large enough amplitude experience larger transverse drag and virtual mass forces when departing from the wall than when returning to it. With moderately oblate bubbles, i.e. in an intermediate Bond number range, this effect is large enough to counteract the repulsive interaction force, forcing such bubbles to perform a periodic zigzagging-like motion at a constant distance from the wall.
\end{abstract}
	
	
\section{Introduction}\label{sec:intro}
In the first part of this investigation (\citealp*{2024_Shi_b}, hereinafter referred to as Part 1), we analyzed the results of a series of simulations revealing the mechanisms governing the lateral migration of freely-deformable gas bubbles rising near a vertical wall in a liquid at rest. The physical parameters were selected so that bubbles rose at moderate Reynolds number and underwent low-to-moderate deformation, ensuring that they would rise in a straight line in the absence of the wall. However, millimetre-size gas bubbles rising in weakly or moderately viscous liquids, especially water, are subject to path instability \citep{1995_Duineveld, 2001_Vries, 2008_Zenit, 2016_Cano-Lozano, 2023_Bonnefis, 2024_Bonnefis}. Consequently, they usually follow either planar zigzagging or (possibly flattened) spiralling paths, both of which exhibit large-amplitude horizontal excursions. Only small enough bubbles can still rise in a straight line when the Reynolds number exceeds a few hundred. These highly inertial regimes, with or without the presence of path instability, are those on which this second part of our investigation focuses.

As discussed in Part 1, interactions between isolated rising bubbles and a vertical wall in moderately inertial regimes are largely governed by two distinct mechanisms. These are the \emph{attractive} inviscid Bernoulli mechanism, predicted by potential flow theory and resulting from the acceleration of the flow in the gap \citep{1976_Wijngaarden, 1977_Miloh, 1993_Kok}, and the \emph{repulsive} vortical mechanism associated with the small flow correction induced by the interaction of the wake with the wall at large distances downstream of the bubble \citep{2002_Takemura, 2003_Takemura, 2015_Sugioka, 2020_Shi, 2024_Shi_a}. The type of near-wall bubble motion depends largely on the relative magnitudes of the irrotational and vortical interaction mechanisms, which in turn depend on the rise Reynolds number, $Re$, and the geometrical aspect ratio, $\upchi$, of the bubble. Here, $Re$ is based on the bubble's equivalent diameter and rise speed, and $\upchi$ is the length ratio of the major to minor axes of the bubble. Provided that $\Rey$ is smaller than a critical value, $\Rey_1$, increasing from $\approx35$ at $\upchi=1$ to $\approx100$ at $\upchi\approx1.5$, the repulsive vortical mechanism dominates, causing bubbles to consistently migrate away from the wall. On the other hand, for $\Rey > \Rey_1$, both mechanisms remain active, and the bubble is first attracted to the wall down to a certain critical distance at which the total transverse force vanishes, and then undergoes either regular or damped transverse oscillations.

In Part 1, the bubble Reynolds number was kept below approximately $200$, and the aspect ratio was, in most cases, smaller than $2$. The aim in this second part is to explore the regime in which bubbles rise with Reynolds numbers of $\mathcal{O}(10^2-10^3)$, while their aspect ratios may vary from $1$ to nearly $3$. In such highly inertial regimes, isolated bubbles rising in an unbounded expanse of a weakly viscous liquid are known to undergo a path instability when their aspect ratio exceeds a critical value $\upchi_c\approx2.0$ \citep{1995_Duineveld, 2001_Vries, veldhuis2007leonardo, 2008_Zenit,2024_Bonnefis}. Depending on the carrying fluid, the corresponding threshold Reynolds number may vary by more than one order of magnitude, being about $670$ for water and about $110$ for a silicone oil five times more viscous than water. Hence, depending on $\upchi$, two sub-regimes exist for Reynolds numbers of $\mathcal{O}(100-1000)$: a stable one, in which the bubble undergoes a moderate deformation and maintains a vertical path in the absence of the wall, and an unstable one, in which its oblateness is somewhat larger and leads to an unstable path. 

The only investigation to date in the first sub-regime appears to be the experiments by \citet{2001_Vries}, some of which were described by \cite{2002_Vries}. There, bubbles rising near a vertical wall in water were observed to undergo a regular bouncing motion when their equivalent radius, $R$, exceeded about $0.4\,\text{mm}$ (corresponding to a Reynolds number in the absence of the wall, $\Rey_\infty$, of approximately $150$). As the bubble size increased further, the amplitude of the transverse oscillations grew. Then, provided $R$ exceeded a second critical value of about $0.6\,\text{mm}$ (corresponding to $\Rey_\infty \approx 370$), the bubble, after colliding with the wall, was able to reach a large wall-normal separation and never returned to the wall. Nevertheless, in the initial stages, these `escaping' bubbles behaved essentially as regular bouncing bubbles, and managed to escape from the wall only after (at least) one period of near-wall bouncing. Clearly, the mechanism promoting the final escape cannot be interpreted as the dominance of the repulsive vortical mechanism summarized above as, if this were the case, the bubble would migrate away from the wall from the very beginning of the interaction sequence. This situation leads to the first objective of this work, which is to clarify the mechanism triggering the escape from the wall of moderately deformed bubbles rising at sufficiently high $\Rey$.\\
\indent On the other hand, bubbles with $\upchi > \upchi_c$ follow an unstable path even in the absence of the wall. For $\upchi\lesssim\upchi_c$, results from Part 1 indicate that the vortical mechanism dominates, causing all bubbles with $\Rey \approx 100$ to consistently migrate away from the wall.
Hence, the repulsive vortical mechanism governs the bubble-wall interaction slightly below the onset of path instability. Moreover, the magnitude of the vorticity generated at the bubble surface (hence the intensity of the wake-wall interaction) increases with increasing $\upchi$ \citep{2007_Magnaudet}. Given these two arguments, one would expect the dominance of the vortical repulsive mechanism to persist as $\upchi$ increases beyond $\upchi_c$. If so, all bubbles undergoing path instability would exhibit a net migration away from the wall, on which path oscillations would superimpose. This behaviour has indeed been reported in previous numerical simulations \citep{2020_Zhang, 2022_Yan, mundhra2023effect} and in the recent experiments by \citet{2024_Estepa-Cantero}, where bubbles were found to follow either planar zigzagging or flattened spiralling paths while gradually migrating away from the wall. The same behaviour was observed in Part 1 with a bubble with $(\upchi,\Rey) \approx (2.1,96)$. In these investigations, highly viscous fluids were considered, so that the Reynolds number remained between $100$ and $200$. In contrast, several experimental studies performed in water \citep{2001_Vries, 2015_Jeong, 2017_Lee, 2023_Cai} considered much larger Reynolds numbers in the range $[700-1100]$, corresponding to equivalent bubble radii from 0.97 to $1.96\,\text{mm}$. The observed paths differed dramatically from those described above: instead of gradually migrating towards the bulk, these high-$\Rey$ bubbles were found to be trapped by the wall, undergoing a zigzagging near-wall motion throughout their ascent. This leads to the second question we wish to examine here, namely the underlying mechanism responsible for the high-$\Rey$ wall-ward trapping of bubbles. Is it linked to the attractive inviscid Bernoulli mechanism, as in the moderately inertial regimes, or is it specifically related to the interaction between the wall and the double-threaded wake that accompanies zigzagging and spiralling bubbles? \\
\indent To make progress on the above questions, we carry out a series of high-resolution simulations covering a significant range of hydrodynamic conditions and analyze the different evolution scenarios. In \S\,\ref{sec:problem_state}, we formulate the problem, specify the range of parameters we consider, and summarize the numerical approach (a series of tests aimed at confirming the adequacy of the grid resolution are detailed in appendix~\ref{sec:app_pre}). Section\,\ref{sec:overview} provides an overview of the observed scenarios, highlighting the existence of several distinct regimes depending on whether the bubble path is stable or not. The escape scenario observed in the absence of path instability is discussed in \S\,\ref{sec:bte}, while those found in the unstable path regime, with or without wall-ward trapping, are discussed in \S\,\ref{sec:zigzag}. Section\,\ref{sec:conclusion} summarizes the main findings of both parts of this investigation.

\section{Statement of the problem and outline of the numerical approach}
\label{sec:problem_state}
\begin{figure}
\centerline{\includegraphics[scale=0.65]{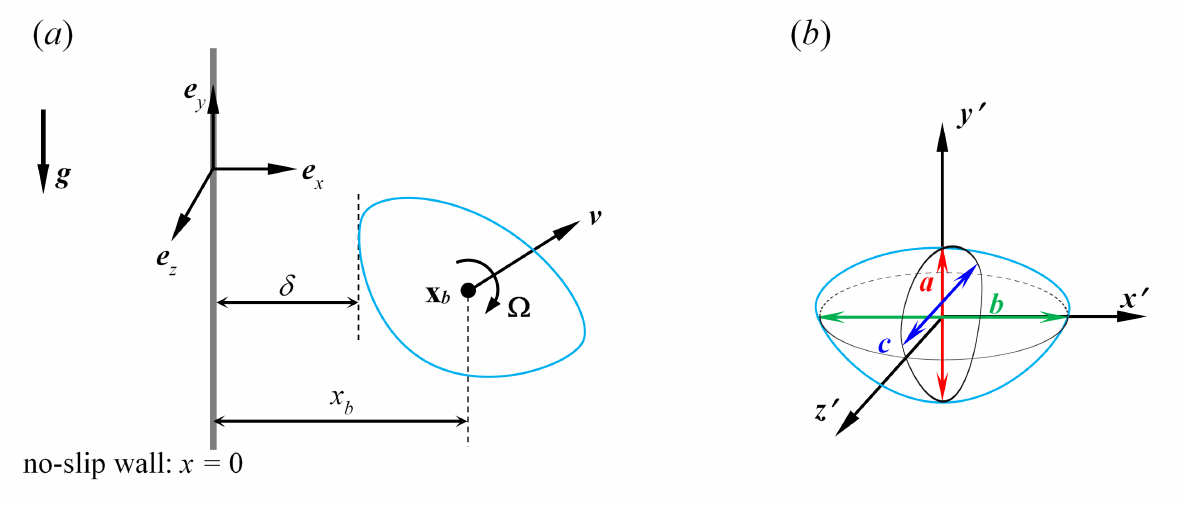}}
\caption{Sketch of the problem. $(a)$: flow configuration and basic quantities characterizing the bubble motion; $(b)$: bubble geometry.}
\label{fig:problem-state}
\end{figure}

An initially spherical gas bubble with radius $R$ rises under the effect of buoyancy in a stagnant liquid in the presence of a nearby vertical wall which we assume to be hydrophilic. Figure \ref{fig:problem-state}$(a)$
specifies the coordinate system, in which the wall lies in the plane $x=0$. The initial and current positions of the bubble centroid are ${\bf{x}}_{b0}=(x_0, 0, 0)$  and ${\bf{x}}_b(t)=(x_b(t), y_b(t), z_b(t))$, respectively, and the minimum gap between the wall and the bubble surface is $\delta(t)$. 

\vspace{-1mm}
The bubble translational velocity is denoted as $\boldsymbol{v}(t)$, and its possible oscillation frequency is $f$. Similarly, the local fluid velocity is $\boldsymbol{u}({\bf{x}},t)$, the vorticity is $\boldsymbol\omega({\bf{x}},t)=\nabla\times\boldsymbol{u}$, and the possible spinning rate (to be defined later) of the interface is $\boldsymbol{\Omega}(t)$. Assuming that the gas-to-liquid density and viscosity ratios keep very small values, and that the initial dimensionless separation $X_0=x_0/R$ is kept fixed (hereinafter $X_0=2$, except in some runs examined in \S\,\ref{sec:bte} and in appendix \ref{sec:app_ini_sep} where the influence of $X_0$ is investigated), the flow and bubble dynamics may entirely be characterized by the Galilei and Bond numbers, respectively defined as
\begin{equation}
Ga=\rho_l g^{1/2} R^{3/2}/{\mu_l}, \quad
Bo=\rho_l g R^2/{\gamma}\,,
\label{eq:ga-bo}
\end{equation}
where $\rho_l$ and $\mu_l$ are the density and dynamic viscosity of the carrying liquid, and $\gamma$ denotes the surface tension. One of the above two control parameters may be replaced with the Morton number $Mo = g \mu_l^4 / (\rho_l \gamma^3) = Bo^3 / Ga^4$, which entirely characterizes the carrying liquid in a given gravitational environment. Once ${\boldsymbol{v}}(t)$ and $f$ are known, the flow dynamics may be characterized by the instantaneous bubble Reynolds number $\Rey(t)=2\rho_l ||\boldsymbol{v}(t)|| R/{\mu_l}$ and the Strouhal number (or reduced frequency) $St=2fR/V_m$, where $V_m$ denotes the time-averaged rise speed. 

In Part 1, we focused on the parameter range $10 \leq Ga \leq 30$ and $0.01 \leq Bo \leq 1.0$, which yielded terminal Reynolds numbers $25 \lesssim \Rey \lesssim 200$ and aspect ratios $1.01 \lesssim \upchi \lesssim 2.1$. Here, we consider the more inertial range $30 < Ga \leq 90$ and $0.02 \leq Bo \leq 2$ in which strongly deformed bubbles with $\upchi \gtrsim 2.0$ rising in an unbounded fluid follow zigzagging or (possibly flattened) spiralling paths  \citep{2008_Zenit, 2016_Cano-Lozano}. 
The corresponding Morton numbers range from $1.2 \times 10^{-13}$ to $1.0 \times 10^{-5}$, i.e. from Galinstan (a liquid metal) to silicone oil DMS-T11 whose kinematic viscosity is ten times that of water at $20^\circ$C. 

\comm{In experimental investigations under a given gravitational environment, the selected gas-liquid system may be characterized through the Morton number $Mo = g \mu_l^4 / (\rho_l \gamma^3) = Bo^3 / Ga^4$. The ranges of $Bo$ and $Ga$ considered in this work correspond to Morton numbers ranging from $1.2 \times 10^{-13}$ to $1.0 \times 10^{-5}$. To connect the $(Bo, Ga)$ sets discussed later with actual gas-liquid systems, table \ref{tab:mo} summarizes the Bond number and bubble size of eight fluids with $Mo$ ranging from approximately $1 \times 10^{-13}$ to $1 \times 10^{-5}$ at the two extremities of the considered range of Galilei number, $Ga = 35$ and $Ga = 90$. The corresponding $R$, $Bo$, and $Ga$ at the threshold of the path instability obtained from linear stability analysis on freely rising deformable bubbles (hereafter abbreviated as LSA) \citep{2024_Bonnefis} are listed as well. Details on the physical properties of the liquids can be found in Part 1 and in prior works \citep{2024_Bonnefis, 2024_Estepa-Cantero}.
}

The results to be discussed below were obtained by solving the three-dimensional time-dependent two-phase Navier-Stokes equations using the open-source flow solver \emph{Basilisk} \citep{2009_Popinet, 2015_Popinet}. Characteristics of this code, averaging rules used to compute the local density and dynamic viscosity of the fluid medium, and conditions imposed at the various boundaries of the computational domain were discussed in Part 1 and are not duplicated here. 
The computational domain is a cubic box with an edge length $L = 480R$, twice that used in Part 1. This larger size allows us to track the bubble over a sufficiently long vertical distance to examine its behaviour under fully-developed conditions. 
The minimum cell size, $\Delta_{\min}$, is decreased down to $\overline\Delta_{\min}\equiv \Delta_{\min}/R \approx 1/68$ close to the interface, and to $\overline\Delta_{\min}\equiv1/34$ in the far wake. Following Part 1, the former minimum is further decreased down to $\approx 1/136$ when $\overline\delta\equiv \delta/R\leq0.15$ to properly resolve the flow in the gap when the bubble gets very close to the wall. The resolution in the far wake is about twice as fine as in Part 1, allowing us to track the details of the wake structure over distances of $O(10R)$ downstream of the bubble. The adequacy of the grid resolution is confirmed through a grid-independence study detailed in appendix \ref{sec:app_pre}. Comparisons between present predictions and experimental results concerning bubbles with non-straight paths rising either in the presence \citep{2024_Estepa-Cantero} or in the absence \citep{1995_Duineveld, 2014_Tagawa} of a wall are also discussed in this appendix. The good agreement obtained with these experimental results confirms the reliability and accuracy of the numerical approach. However, it must be stressed that even the maximum refinement $\overline\Delta_{\min}\equiv1/136$ is not sufficient to properly resolve the flow in the gap when $\overline\delta(t)$ becomes extremely small. In such cases, lubrication effects in the gap are not fully captured, leading to what is referred to as a `bubble-wall collision' in the next sections. It was shown in Part 1 that this under-resolution has virtually no effect on the bouncing frequency and only lowers the maximum separation achieved by the bubble after a `collision' event by a few percent.\\
\indent To roughly characterize the bubble geometry, we need to define the orientation and length $a$ of the minor axis, and the lengths $b$ and $c$ of the major axes in the wall-normal and wall-parallel planes, respectively. To this end, we first consider the wall-normal plane containing the bubble centroid and identify the shortest and longest axes passing through this centroid, following \cite{2021_Zhang}. With this definition, there is in general no reason for these two axes (denoted as $x'$ and $y'$ in  figure \ref{fig:problem-state}$(b)$, respectively) to be strictly orthogonal. The third axis, denoted as $z'$ in the figure, is parallel to the wall and passes also through the bubble centroid. Nevertheless, the maximum horizontal extension of the bubble in planes parallel to the wall may not lie along this axis if the bubble exhibits asymmetries. To better approach this maximum extension, we 
 identify the length $c$ as that of the longest horizontal segment connecting two points of the bubble surface and lying in the 
 wall-parallel plane passing through the centroid. We are then in position to compute the principal aspect ratio $\upchi=b/a$, and the equatorial axes ratio, $\upchi_\perp=b/c$. 
In order to obtain a global characterization of the fluid motion at the interface, we also introduce the interface spinning rate, following a suggestion of \cite{rastello2009drag}. Based on the velocity of all fluid elements at the interface, we define this spinning rate, $\boldsymbol{\Omega}$, as 
\begin{equation}
  \boldsymbol{\Omega}(t) = \frac{3}{2\mathcal{V}_s}\int_{\mathcal{V}_s} \frac{\boldsymbol{r}(t) \times \boldsymbol{u}(\boldsymbol{r},t)}{||\boldsymbol{r}(t)||^2}~\mathrm{d} \mathcal{V}_s\,,
\label{eq:spin}
\end{equation}
where $\mathcal{V}_s$ is the volume of the thin film made of the computational cells straddling the gas-liquid interface, i.e. those in which the gas and the liquid are both present, and $\boldsymbol{r}(t)=\mathbf{x} - \mathbf{x}_b(t)$ is the position vector with respect to the bubble centroid. The definition \eqref{eq:spin} may be shown to yield the exact spinning rate of a sphere undergoing a rigid-body rotation about one of its diameters. However, bubbles deform upon time and the carrying fluid obeys a shear-free condition at their surface, making the angular dynamics of bubbles very different from those of usual rigid bodies surrounded by a fluid obeying a no-slip condition. This is why the term `spinning rate' must not be misunderstood, as a bubble with $ \boldsymbol{\Omega}\neq\bf{0}$ may not rotate as a whole. Therefore, in general, $\boldsymbol{\Omega}$ must only be regarded as a three-dimensional measure of the average fluid rotation over the bubble surface.\\
\indent In the following sections, we make extensive use of dimensionless quantities to describe the flow field and bubble motion. To this end, all variables are normalized using $R$ and $\sqrt{R/g}$ as characteristic length and time scales, respectively. The dimensionless time and local position are denoted as $T$ and ${\bf{X}}=(X,Y,Z)$, respectively, while the dimensionless frequency and position of the bubble centroid are $\overline{f}$ and ${\bf{X}}_b=(X_b,Y_b,Z_b)$, respectively. Similarly, in the $({\bf{e}}_x,{\bf{e}}_y, {\bf{e}}_z)$-basis, the $i$-th component of the dimensionless bubble velocity, spinning rate, fluid velocity and vorticity are denoted as $V_i$, $\overline\Omega_i$, $U_i$ and $\overline\omega_i$, respectively.


\section{Overview of the results}\label{sec:overview}
\begin{figure}
\centerline{\includegraphics[scale=0.65]{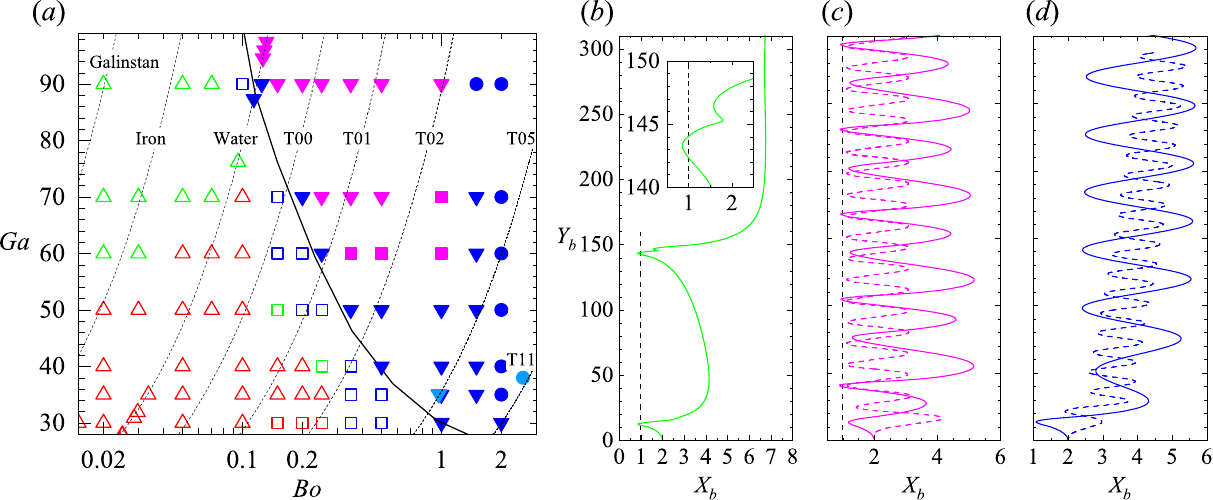}}
\vspace{-1mm}
\caption{Styles of paths observed in the simulations. ($a$): phase diagram in the ($Bo, Ga$)-plane; ($b,c,d$): typical trajectories illustrating the bouncing-tumbling-escaping (BTE) scenario at $(Bo, Ga)=(0.05, 70)$, the near-wall zigzagging (NWZ) motion with (solid line, $(Bo, Ga)=(0.25, 90)$) and without (dashed line, $(Bo, Ga)=(1, 70)$) bubble-wall collisions, and the wavy migration away (WMA) scenario with a planar zigzagging path ($(Bo, Ga)=(0.2, 70)$, solid line; $(Bo, Ga)=(1.5, 50)$, dashed line), respectively. Solid line in ($a$):  neutral curve corresponding to the onset of path instability in an unbounded fluid \citep{2024_Bonnefis}; dashed lines: iso-$Mo$ lines for different liquids 
(see table 1 in Part 1 for their physical characteristics). Open symbols denote cases in which bubbles do not undergo a path instability, with \textcolor{red}{$\triangle$} and \textcolor{red}{$\square$}: periodic near-wall bouncing with and without bubble-wall collisions, respectively; \textcolor{green}{$\square$}: damped bouncing; \textcolor{blue}{$\square$}: migration away from the wall; \textcolor{green}{$\triangle$}: BTE. Solid symbols denote scenarios observed in the presence of path instability, with \textcolor{mymagenta}{$\blacktriangledown$} and \textcolor{mymagenta}{$\blacksquare$}: NWZ with and without bubble-wall collisions, respectively; \textcolor{blue}{$\blacktriangledown$} and {\raisebox{-0.6ex}{\Large{\color{blue}\textbullet}}}: WMA with a planar zigzagging or a (possibly flattened) spiralling path, respectively. 
Data at $Ga\approx76$, $Ga\approx87$ and $Ga\geq94$ in water are taken from experiments by \cite{2001_Vries}, while \textcolor{myblue}{$\blacktriangledown$} (silicone oil DMS-T05 at $Ga=35$), and {\raisebox{-0.6ex}{\Large{\color{myblue}\textbullet}}} (water-glycerol mixture at $Ga\approx38$) are taken from those of \cite{2024_Estepa-Cantero}.  
}
\label{fig:traj_sum}
\end{figure}

\begin{figure}
\centerline{\includegraphics[scale=0.7]{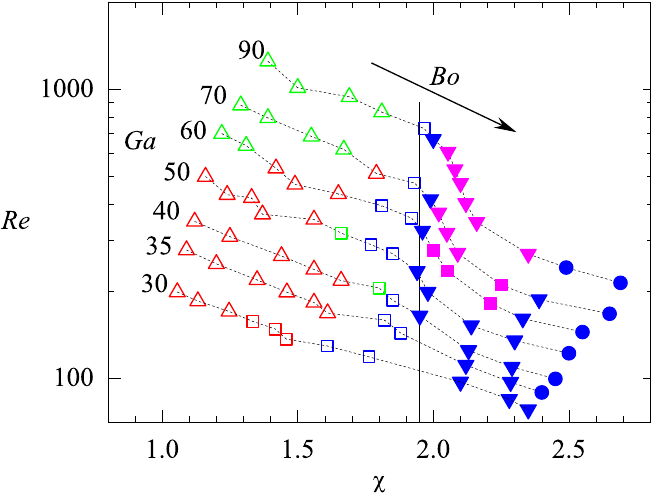}}
\vspace{-1mm}
\caption{Styles of paths in the ($\upchi, \Rey$)-plane. For caption, see figure \ref{fig:traj_sum}. Vertical line: $\upchi=1.95$; thin dashed lines: iso-$Ga$ curves, with $Ga$ increasing from $30$ to $90$ from bottom to top, and $Bo$ increasing from $0.02$ to $2$ from left to right on each iso-$Ga$ curve. For cases below the neutral curve, values of $\Rey$ and $\upchi$ are based on final conditions, except in cases with near-wall oscillations, for which average values taken over a single period are used. For cases beyond the neutral curve, values of $\Rey$ and $\upchi$ correspond to averages taken over one period of the zigzagging or spiralling path in an unbounded fluid.}
\label{fig:chi-vs-re}
\end{figure}

Figure \ref{fig:traj_sum}($a$) summarizes in the form of a phase diagram in the $(Bo, Ga)$ plane the various types of near-wall motion observed in the simulations; results for $Ga = 30$ were taken from Part 1. The solid line crossing the phase diagram corresponds to the neutral curve (reproduced from \citet{2024_Bonnefis}), beyond which the bubble path becomes unstable when the fluid domain is unbounded. Figure \ref{fig:chi-vs-re} is the equivalent of figure \ref{fig:traj_sum}($a$) in the $(\upchi, \Rey)$-plane. 

\subsection{Below the neutral curve}\label{sec:ove_1}
We first review the behaviours observed under conditions where the bubble would rise in a straight line in an unbounded fluid, which corresponds to cases located below the neutral curve in figure \ref{fig:traj_sum}($a$). Up to $Ga = 50$, the three types of motion already analysed in Part 1 are observed. Specifically, periodic near-wall bouncing takes place at low $Bo$, while a regular migration away from the wall is observed at large $Bo$, with, occasionally, damped bouncing motions in between. As discussed in Part 1, the  mechanisms governing the transition between these three styles of path result from the competition between the irrotational and vortical interaction mechanisms. A new type of motion occurs for $Ga \geq 60$, in which  the bubble manages to escape the near-wall region after one or two rounds of near-wall bouncing. Figure \ref{fig:traj_sum}($b$) shows a typical path corresponding to this scenario. After two rounds of bouncing, the bubble briefly migrates towards the wall (as highlighted in the inset of panel ($b$)), then quickly departs from it, eventually reaching a wall-normal position $X_b \approx 6.7$ at a vertical position $Y_b \approx 200$. Afterwards, the bubble almost rests at this wall-normal position, its wall-normal velocity exhibiting only minute values. This is because the disturbance induced by a bubble moving at $\Rey = O(100)$ decays essentially as the inverse of the cube of the distance to its centroid, following the prediction of potential flow theory. Hence, the presence of the wall is only weakly `felt' by the bubble in the late stages of its ascent, so that it rises almost vertically as in an unbounded fluid. This path evolution differs from that observed in the damped bouncing regime, where the bubble remains close to the wall throughout its ascent and finally rests at a wall-normal position where the total transverse force vanishes.

This escape scenario specific to highly inertial regimes, particularly the brief wall-ward migration preceding the escape, was observed experimentally in water by \citet{2002_Vries} (see figure 5 therein) with an air bubble corresponding to $(Bo, Ga) \approx (0.095, 75.5)$. According to figure \ref{fig:traj_sum}$(a)$, this scenario only occurs for nearly-spherical bubbles at $Ga = 60$, but quickly dominates the entire region $Bo < 0.1$ when $Ga$ is increased beyond. Figure \ref{fig:chi-vs-re} makes it clear that this scenario is encountered for $\Rey \gtrsim 600$ and $\upchi \lesssim 1.85$. The underlying mechanisms will be discussed in \S\, \ref{sec:bte}. Here, we just point out that, prior to the escape, a strong vortical layer forms in the vicinity of the bubble surface upon its collision with the wall, causing the $z$-component of the spinning rate, $\overline\Omega_z$, to take large values. This spinning motion yields a sizeable Magnus-like force pointing away from the wall, which promotes the bubble escape. In what follows, we refer to this type of path evolution as the bouncing-tumbling-escaping (BTE) scenario.

\begin{figure}
\centerline{\includegraphics[scale=0.65]{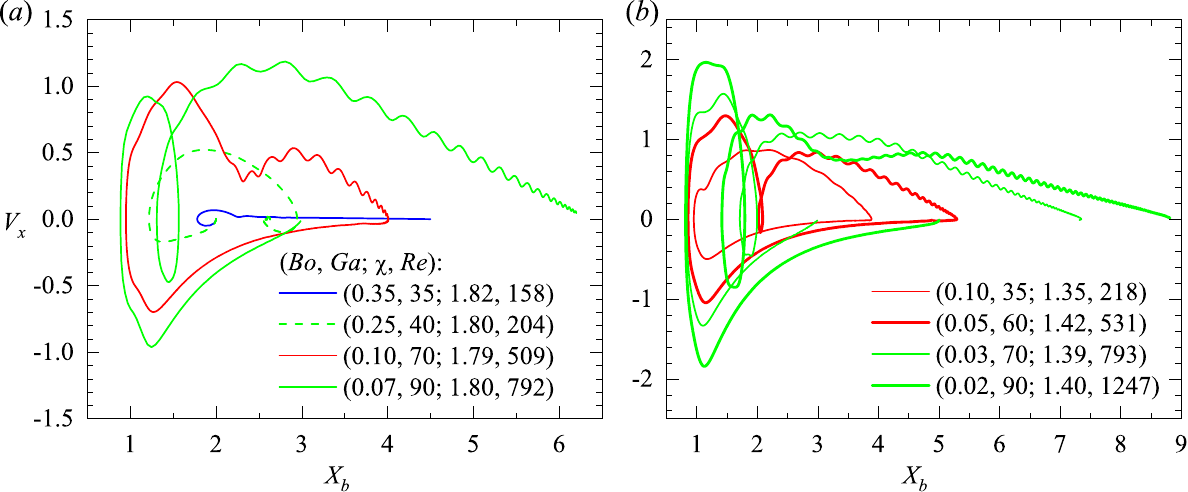}}
\vspace{-1mm}
\caption{Variations of the bubble wall-normal velocity, $V_x$, as a function of the bubble-wall distance, $X_b$, for selected cases at ($a$): $\upchi \approx 1.8$, and ($b$): $\upchi \approx 1.4$. The corresponding values of $Bo$, $Ga$, and the resulting $\Rey$ are indicated in each panel. The initial position is set as $(X_0, V_x) = (2, 0)$. 
In periodic bouncing cases (red lines), only the variation during the fully-developed stage is shown. In BTE cases (solid green lines), only variations during the last cycle of motion, starting at a time when the bubble begins to migrate towards the wall, are shown.}
\label{fig:u-vs-x-1}
\end{figure}
To illustrate the influence of the Reynolds number on the style of path, we select a series of results obtained at two specific values of the bubble aspect ratio, $\upchi=1.4$ and $\upchi=1.8$. Figure \ref{fig:u-vs-x-1} shows the variations of the wall-normal velocity, $V_x(T)$, with the lateral position, $X_b(T)$, as $\Rey$ increases by one order of magnitude, from values of $\mathcal{O}(10^2)$ to $\mathcal{O}(10^3)$. For $\upchi \approx 1.8$, the bubble migrates away from the wall at $\Rey = 158$, experiences damped near-wall oscillations at $\Rey = 204$, and bounces periodically on the wall at $\Rey = 509$. In these two bouncing cases, the maximum lateral separation between the bubble and wall increases from $3$ to $4$ as $\Rey$ increases. At $\Rey = 792$, the bubble manages to reach a lateral position $X_c = 6.3$, where it rests. The brief wall-ward migration taking place before the escape corresponds to the short period of time when $V_x(T) < 0$ in the loop formed by the corresponding curve. For $\upchi = 1.4$, the bubble already experiences periodic near-wall bounces at $Ga = 35$, corresponding to $\Rey = 218$. At $\Rey = 531$, while it still undergoes regular bouncing, its lateral motion weakly reverses during the departing stage. The maximum lateral position is $X_b=5.3$, and this large separation results in a very slow bouncing frequency, $St = 0.0055$. 
As $\Rey$ increases further, the bubble escapes from the wall, resting eventually at a lateral position $X_c = 7.3$ at $\Rey = 793$ and $X_c=8.8$ at $\Rey = 1247$.

Figure \ref{fig:u-vs-x-1} also allows the effects of the aspect ratio to be appreciated. First, it is seen that in the two cases with $\Rey$ close to $200$, the less deformed bubble experiences periodic bounces, while that with $\upchi=1.80$ follows a damped bouncing evolution. In the two cases with $\Rey$ close to 500, both bubbles undergo regular bouncing but their maximum lateral position reduces from 5.3 at $\upchi = 1.42$ to $4.0$ at $\upchi = 1.79$. Accordingly, the reduced frequency increases from $0.0055$ at $\upchi = 1.42$ to $0.013$ at $\upchi = 1.79$. Last, when $\Rey \approx 790$, both bubbles manage to escape from the near-wall region and their final wall-normal position also decreases as $\upchi$ increases. 

\subsection{Beyond the neutral curve}\label{ove_2}
\begin{figure}
\centerline{\includegraphics[scale=1]{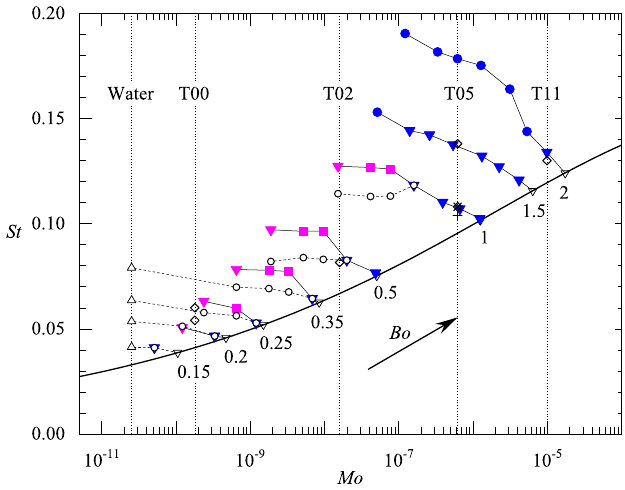}}
\vspace{-1mm}
\caption{Variation of the Strouhal number, $St$, of path oscillations as a function of the Morton number. Solid symbols: present results in the wall-bounded configuration (for caption, see figure \ref{fig:traj_sum}); open circles: present  results in an unbounded fluid. Thin solid and dashed lines connect $St$ values corresponding to a fixed $Bo$ and different $Ga$ (hence, $Mo$) in the wall-bounded and unbounded configurations, respectively. Thick solid line: neutral curve of path instability in an unbounded fluid \citep{2024_Bonnefis}; $\triangledown$: selected values of the critical Strouhal number, $St_c$, from the same reference
; $\triangle$: experimental data in water \citep{1994_Duineveld, 1995_Duineveld, veldhuis2007leonardo, 2008_Veldhuis, 2015_Jeong} (values were interpolated from neighbouring $Bo$). $+$ and $\times$: data at $Bo \approx 1$ from experiments in silicone oil DMS-T05 by \cite{2009_Zenit_a} and \citet{2024_Estepa-Cantero}, respectively; $\diamond$: numerical data for various liquids in an unbounded fluid \citep{2016_Cano-Lozano}. }
\label{fig:f-zigzag}
\end{figure}
In figure \ref{fig:chi-vs-re}, bubbles with an aspect ratio larger than $\approx1.95$ are found to follow zigzagging or spiralling paths, consistent with the experimental observations of \citet{2008_Zenit}. Similarly, in figure \ref{fig:traj_sum}$(a)$, path instability is seen to take place for all $(Bo, Ga)$ sets located beyond the neutral curve determined for bubbles rising in an unbounded fluid. This observation suggests that the presence of the wall plays no role in the occurrence of path instability. However, in its early stage, it selects the orientation of the plane in which the bubble oscillates, making path instability arise through an imperfect bifurcation. 
Up to $Ga=50$, all bubbles experiencing path instability migrate away from the wall. We refer to this type of motion as the wavy migration-away (WMA) scenario. In this regime, bubbles maintain a zigzagging path in the wall-normal plane up to $Bo \approx1.5$ (see the examples in figure \ref{fig:traj_sum}$(d)$). Then, this path transitions to a spiralling motion at larger Bond numbers. Bubbles following the WMA scenario were observed experimentally by \citet{2024_Estepa-Cantero}. 
The corresponding data, shown in figure \ref{fig:traj_sum}$(a)$, corroborate present findings. For $Ga>50$, the same scenario is encountered both for $Bo > 1$ and within a narrow band of significantly lower Bond numbers lying just above the neutral curve, e.g. $Bo\approx0.2$ for $Ga=70$ and $Bo\approx0.12$ for $Ga=90$.\\
\indent 
When $Ga$ is larger than $50$ and $Bo$ is above the aforementioned band but below unity, a new type of motion emerges. In this intermediate range, the bubble is trapped near the wall, ultimately undergoing a near-wall zigzagging-like (NWZ) motion. Bubble-wall collisions due to an incomplete resolution of lubrication effects in the immediate vicinity of the wall (see \S\,\ref{sec:problem_state}) occur for $Ga \gtrsim 70$. Two bubble paths typical of this regime, with and without bubble-wall collisions, are shown in figure \ref{fig:traj_sum}$(c)$. Although the upper limit of present simulations is $Ga=90$, available data indicate that this scenario still holds at larger $Ga$. In particular, it was identified in experiments performed in pure water \citep{2001_Vries, 2015_Jeong, 2017_Lee, 2023_Cai} with millimetre-sized air bubbles corresponding to $95\lesssim Ga\lesssim270$ and $0.125\lesssim Bo\lesssim0.5$. It is important to stress that the NWZ regime and the near-wall bouncing regime observed for low enough $Bo$ and $Ga$ below the neutral curve (open red symbols in figure \ref{fig:traj_sum}$(a)$) are totally distinct. Indeed, although bubbles follow periodic near-wall paths in both cases, the mechanisms that govern the two regimes differ drastically, as will become clear during the analysis of the NWZ regime in \S\S\,\ref{sec:nwz}-\ref{sec:nwz_mec}. \\
\indent Figure \ref{fig:f-zigzag} shows how the Strouhal number (or reduced frequency), $St$, of the transverse oscillations vary from one fluid to the other. In the simulations, variations of $Mo$ are achieved by varying $Ga$ while keeping $Bo$ constant. 
For each iso-$Bo$ series (thin solid and dashed lines), $St$ is seen to decrease with increasing $Mo$ (hence, decreasing $Ga$). The reduced frequency approaches the threshold value, $St_c$, predicted by linear stability analysis \citep{2024_Bonnefis} at the maximum $Mo$ at which path instability takes place at the considered $Bo$. This finding indicates that the presence of the wall does not affect the oscillation frequency of bubbles close to the threshold of path instability. 
To check whether or not this is still the case further away from the threshold, we ran additional runs without the presence of the wall for $(Bo,Ga)$ sets for which the NWZ evolution is observed when the wall is present. Comparing both sets of results reveals that interactions with the wall increase the reduced frequency for $Bo\geq0.25$. For instance, with $Bo=0.5$ and $Mo = 1.9 \times 10^{-9}$, $St$ increases by nearly $20\%$ from the unbounded to the wall-bounded configuration. Some available experimental and numerical data, all of which were obtained in an unbounded fluid except those of \cite{2024_Estepa-Cantero}, are also included in the figure. These data are found to agree well with present predictions, confirming the accuracy of our simulations in the unbounded configuration (the low-$Mo$ simulations of \cite{2016_Cano-Lozano} suffer from under-resolution, which is why the data reported for $Mo=\mathcal{O}(10^{-10})$ slightly deviate from present predictions).\\
\indent In the following sections, we examine in more detail the results corresponding to the three regimes that were not observed in Part 1, namely the BTE scenario which takes place at large enough $Ga$ in the stable path regime, and the WMA and NWZ scenarios, which are both observed beyond the threshold of path instability. 

\section{BTE regime}
\label{sec:bte}
\begin{figure}
\centerline{\includegraphics[scale=0.65]{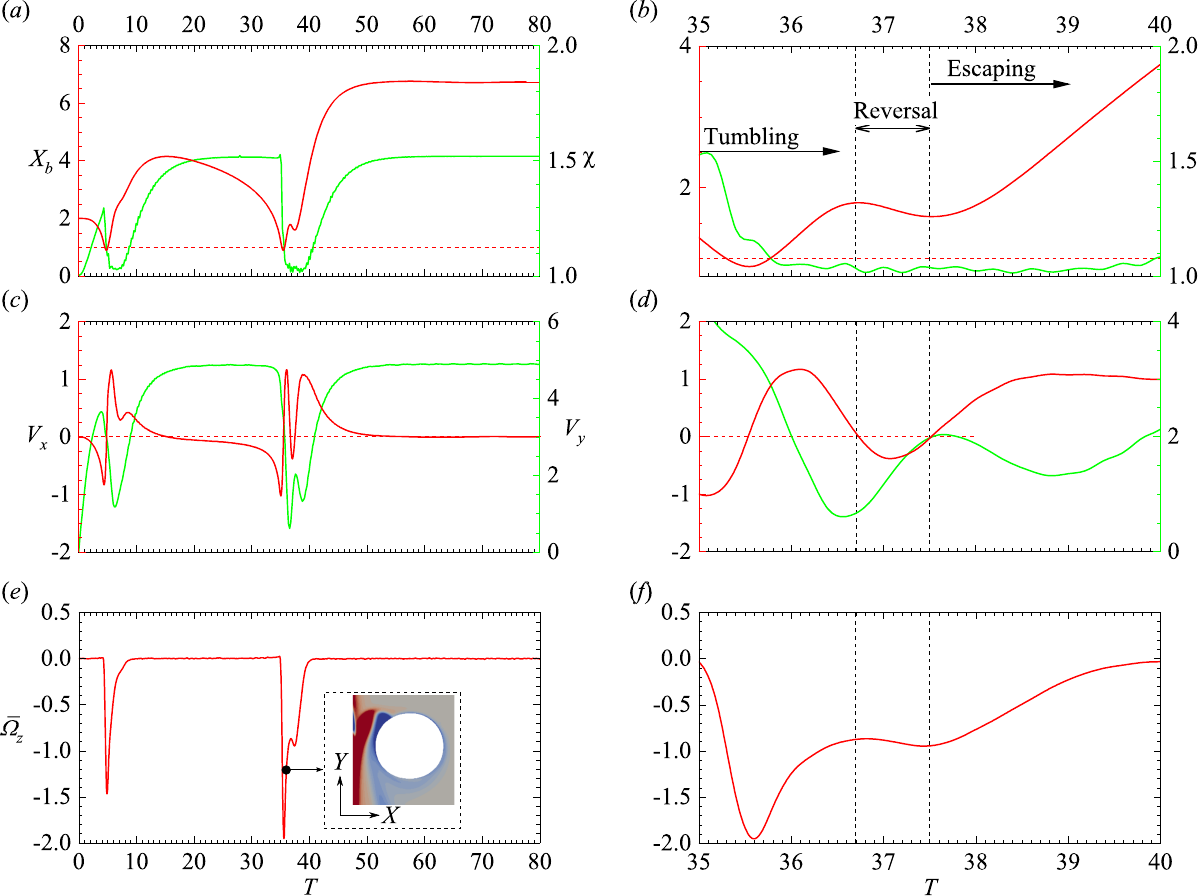}}
\vspace{-1mm}
\caption{Evolution of several characteristics of the bubble motion during a BTE scenario for $(Bo, Ga) = (0.05, 70)$. The right panels provide a zoom of the evolution shown in the left panels in the time interval $35 \leq T \leq 40$. ($a$) and ($b$): wall-normal bubble position (red line and left axis) and bubble aspect ratio (green line and right axis). ($c$) and ($d$): wall-normal (red line and left axis) and vertical (green line and right axis) velocities of the bubble centroid. ($e$) and ($f$): spinning rate of the bubble surface. The inset in panel ($e$) displays the iso-contours of the spanwise vorticity $\overline{\omega}_z$ at $T=36$ in the symmetry plane $Z=Z_b$; red and blue colours refer to positive and negative values, respectively, with a maximum magnitude of $5$.}
\label{fig:tre_motion}
\end{figure}

Figure \ref{fig:tre_motion} illustrates the evolution of several indicators for a bubble with $(Bo, Ga) = (0.05, 70)$ following a BTE evolution. 
According to panel ($a$), the bubble bounces twice before escaping from the near-wall region. Before each bounce, it first collides with the wall, as indicated by values of $X_b$ (the distance from the bubble centroid to the wall) smaller than one. Upon collision, the aspect ratio quickly decreases to a minimum close to $1.0$ and the bubble keeps this virtually spherical shape for about $3-4$ time units (see the detailed evolution in panel ($b$)). 
Panels ($c,d$) show that, upon the second collision, the bubble rise velocity, $V_y$, falls dramatically by a factor of six, reducing from $4.8$ at $T=33$ to $0.75$ at $T=36.5$. Meanwhile, the direction of the wall-normal velocity, $V_x$, reverses three times. Between the last two reversals, $V_x$ is negative, so that the bubble undergoes a short wall-ward migration prior to escaping definitively away from the wall region (mentioned as the `Reversal' period in panel ($b$)). 

Examination of the flow field in the vicinity of the bubble helps elucidate key aspects of the mechanisms involved. The inset in figure \ref{fig:tre_motion}($e$) shows the iso-contours of the spanwise vorticity component, $\overline{\omega}_z$, in the wall-normal symmetry plane $Z=Z_b$ just after the collision ($T=36$). This component is seen to be uniformly negative in the bubble vicinity, indicating that fluid particles experience a clockwise rotation about the $Z$-axis. The global rotation of fluid particles at the bubble surface may be quantified using the spinning rate $\boldsymbol{\overline{\Omega}}$ defined by \eqref{eq:spin}. 
The evolution of its spanwise component, $\overline{\mathit{\Omega}}_z$, is illustrated in figures \ref{fig:tre_motion}($e,f$). Upon the first collision, $|\overline{\mathit{\Omega}}_z|$ quickly increases from zero to $\approx1.4$. Combined with the simultaneous sharp decrease of $V_y$, this suggests that part of the fluid kinetic energy associated with the bubble translation is converted into `rotational' kinetic energy. The fluid surrounding the bubble being spinning,  the bubble experiences a lift force $\boldsymbol{F}_L^\text{M}\propto\boldsymbol{\overline{\Omega}}\times \boldsymbol{V}$, which may be interpreted as a Magnus-like force. As \,$\overline{\mathit{\Omega}}_z$, $V_x$ and $V_y$ are all of order unity during this transient stage, this force is expected to be comparable in magnitude to the buoyancy force. Since $V_y$ is always positive, the wall-normal component of this force is positive, helping the bubble move away from the wall. This is the key mechanism that makes the BTE scenario possible. An essential characteristics of the Magnus-like force is that  it keeps significant values over a long post-collision period of time, while the bubble has already moved a substantial distance away from the wall. For instance, figure \ref{fig:tre_motion}($b$) indicates that the gap is already one bubble radius wide $(X_b=2)$ at $T=38.3$, a moment at which, according to panels $(d)$ and $(f)$, the wall-normal component of $\boldsymbol{\overline{\Omega}}\times \boldsymbol{V}$ is still close to unity. \\
\indent When $V_x$ is positive, the vertical component $\boldsymbol{F}_L^\text{M}\cdot{\bf{e}}_y$ is negative, thus counteracting buoyancy. This leads to a sharp decrease in $V_y$ during the time interval $35.5 \leq T \leq 36.7$, and to a lesser extent from $T=37.5$ to $T =39$. Conversely, during the short wall-ward stage noticed for $36.7\leq T\leq37.5$, this component cooperates with the buoyancy force, resulting in an increase of the rise speed. Of course, viscous effects and liquid inertia play a central role beyond the transient stage during the Magnus-like force controls the bubble dynamics. Since near-wall dissipation increases as the gap shrinks, viscous processes are responsible for the sharp decrease of the rise speed prior to the collision $(33.5\leq T\leq35.5$). Conversely, inertial effects associated with the amount of liquid displaced by the bubble result in an added-mass force that limits the time variations of $V_y$. This is why the rise speed is still only half its pre-collision value at $T=40$, while $\overline{\mathit{\Omega}}_z$ has already almost returned to zero. 

\begin{figure}
\centerline{\includegraphics[scale=0.65]{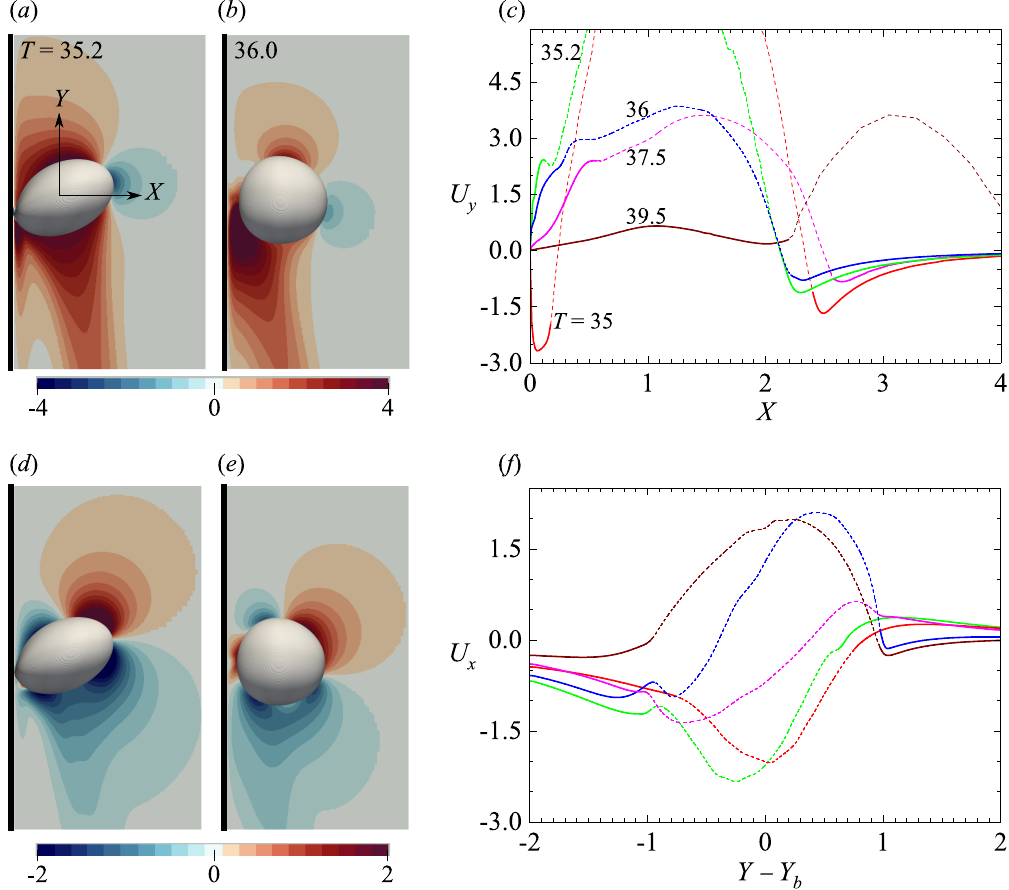}}
\vspace{-1mm}
\caption{Evolution of the vertical (top row) and wall-normal (bottom row) fluid velocity in the vicinity of a bubble with $(Bo, Ga) = (0.05, 70)$. ($a-b$) and ($d-e$): distribution in the wall-normal symmetry plane $Z = Z_b$ at $T = 35.2$ and $36$, respectively. The bubble is rising upwards, and the wall is shown with a thick black line on the left. ($c$): distribution of $U_y$ along the horizontal line $Y = Y_b, Z = Z_b$, with solid and dashed lines showing the velocity outside and inside the bubble, respectively. $(d)$: same for $U_x$ along the vertical line $X = X_b, Z = Z_b$. 
}
\label{fig:tre_v_in_gap}
\end{figure}

Figure \ref{fig:tre_v_in_gap} shows the bubble shape and position and the velocity distribution of the surrounding liquid in the symmetry plane $Z = Z_b$ just before and after the second collision. 
Panel $(c)$ reveals that the extremum of $U_y$ in the gap experiences an abrupt change from $-2.6$ at $T=35$ to $\approx2.5$ at $T=35.2$. The maximum $U_y$ goes on increasing after the collision, reaching a value of $4.2$ at $T=36.5$ (not shown), before relaxing slowly to zero as the bubble moves away from the wall. The massive flow reversal in the gap at the beginning of the sequence results from the rapid decrease of the bubble rise speed, which forces fluid elements initially located in the wake to catch up with those located on both sides of the bubble and eventually replace them. The consequences of this catch-up process are milder along the fluid-facing side of the bubble, where the negative vertical velocity is significantly reduced but does not change sign over an $\mathcal{O}(1)$-thick fluid layer. 
The positive $U_y$ in the gap and their negative counterpart on the fluid-facing side both provide negative contributions to the $z$-component of the cross product $({\bf{X}}-{\bf{X}}_b)\times {\bf{U}}$ at the bubble surface, hence to $\overline\Omega_z$. However, the contribution coming from the gap is $3-4$ times larger, and the difference is even larger before the collision, owing to the weak negative values of $U_y$ on the fluid-facing side ($U_y\approx0$ at the position where the solid green line corresponding to $T=35.2$ turns dashed). Owing to the predominant suction of the wake towards the gap region, the wall-normal velocity is negative at the back of the bubble, as panels $(d-e)$ confirm. These negative $U_x$ values are significantly larger than their positive counterpart ahead of the bubble (see panel $(f)$), providing another sizeable negative contribution to $\overline\Omega_z$.  

\begin{figure}
\centerline{\includegraphics[scale=0.7]{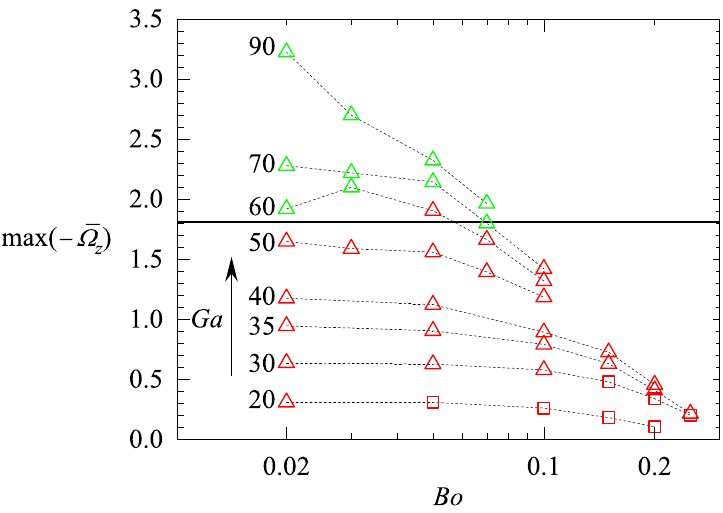}}
\vspace{-1mm}
\caption{Variation of the maximum spinning rate with the Bond and Galilei numbers. Symbols are identical to those in figure \ref{fig:traj_sum}. 
The thick horizontal line corresponds to $\max{(|\overline{\mathit{\Omega}}_z|}) = 1.8$.}
\label{fig:omegaz_sum}
\end{figure}
The BTE regime may set in only if the spinning rate resulting from the mechanism described above is strong enough. Figure \ref{fig:omegaz_sum} displays the maximum spinning rate computed along the bubble path for all cases corresponding to a periodic bouncing or a BTE regime; some cases with $Ga\leq30$ considered in Part 1 are also included. Simulations corresponding to $40 \leq Ga \leq 70$ where run with an initial separation $X_0 = 3.5$, to ensure that the possible escape event takes place after the first collision (see appendix \ref{sec:app_ini_sep} for a discussion of the influence of $X_0$ on the escape process). These results suggest that the bubble path transitions from the periodic bouncing regime to the BTE regime when the maximum spinning rate exceeds a critical value close to $1.8$.\\
\indent To get some more insight into the magnitude of the Magnus-like force, an estimate of the other transverse forces acting on the bubble during its escape from the wall region is required. Some of the kinematic data provided in panels $(b)$,$(d)$ and $(f)$ of figure \ref{fig:tre_motion} may be used for this purpose, noting that the bubble remains nearly spherical in the time interval $36\lesssim T\lesssim40$ as panel $(b)$ indicates. Consider for instance the situation at $T=37.5$, a moment at which the transverse velocity, $V_x$, hence the transverse quasi-steady viscous drag force, is zero. At this moment, $X_b\approx1.6$, $V_y\approx2.0$, $\overline\Omega_z\approx-0.95$. Once normalized by $\rho_l gR^3$, the Magnus-like force may be written in the form $\boldsymbol{F}_L^\text{M}=\frac{4}{3}\pi C_L^\Omega\boldsymbol{\overline{\Omega}}\times \boldsymbol{V}$, with $C_L^\Omega$ being the (unknown) Magnus lift coefficient. Hence, the wall-normal component of this force at this position is approximately $7.95C_L^\Omega$. 
At the same transverse position, the attractive force $\boldsymbol{F}_P=-\frac{\pi}{2}C_PV_y^2{\bf{e}}_x$ resulting from the Bernoulli effect is close to $-0.38$, based on the estimate of the interaction coefficient $C_P(X_b=1.6)\approx0.06$ resulting from equation 7 of \cite{2003_Takemura}. As discussed in Part 1, there is a large uncertainty in the magnitude of the transverse component of the inertia-induced force, owing to the unknown contribution of the trailing vortices entrained laterally by the bubble. With all due caution, an order-of-magnitude estimate of this force may be obtained by neglecting this contribution. With this assumption, the transverse inertia-induced force, $\boldsymbol{F}_I\cdot{\bf{e}}_x=-\frac{4}{3}\pi C_IdV_x/dT$ may be estimated by simply considering that the inertia-induced coefficient, $C_I$, is close to the familiar added-mass coefficient of a sphere accelerating in an unbounded fluid, i.e. $C_I\approx\frac{1}{2}$. With $dV_x/dT\approx1.45$ at $T=37.5$, this estimate yields $\boldsymbol{F}_I\cdot{\bf{e}}_x\approx-3.04$, suggesting that this contribution dominates the overall force that resists the bubble escape.  Summing the two resistive contributions and assuming that the repulsive Magnus-like force is almost in balance with the overall attractive force implies $C_L^\Omega\approx0.43$. A similar estimate may be performed at $T=38.3$, a moment when the gap thickness equals the bubble radius ($X_b=2.0$), so that $C_P\approx0.024$. At this moment, $V_x\approx1.0$, $V_y\approx1.6$, $\overline\Omega_z\approx-0.6$ and $dV_x/dT\approx0.8$, which yields $\boldsymbol{F}_P\cdot{\bf{e}}_x\approx-0.1$ and $\boldsymbol{F}_I\cdot{\bf{e}}_x\approx-1.68$. At this position, the Reynolds number $Re=2Ga(V_x^2+V_y^2)^{1/2}$ is close to $265$, making it possible to estimate the quasi-steady transverse drag force using Moore's high-Reynolds-number prediction \citep{1963_Moore}. In present notations, this yields $\boldsymbol{F}_D\cdot{\bf{e}}_x\approx-12\pi(1-2.2Re^{-1/2})Ga^{-1}V_x\approx-0.47$.  Adding all three attractive contributions together and balancing with the Magnus-like force now implies $C_L^\Omega\approx0.56$. The above two predictions for the Magnus lift coefficient, corroborated by similar estimates for other $(Bo,Ga)$ sets, suggest that  $C_L^\Omega\approx0.5\pm0.07$. Although this result is reminiscent of the well-known inviscid prediction for the lift coefficient $C_L^s$ of a sphere immersed in a weak linear shear flow, $C_L^s=\frac{1}{2}$, this is presumably largely coincidental given the very different nature of the present conditions, in which unsteadiness, flow inhomogeneity and wall vicinity play a central role.

\section{Regimes encountered beyond the path instability threshold}
\label{sec:zigzag}
\subsection{WMA regime}
\label{sec:uma1}
The first phenomenology encountered beyond the neutral curve, at least up to $Ga=90$, corresponds to the WMA regime. To illustrate this regime, we select a bubble with $(Bo, Ga) = (1.5, 50)$, i.e. a Morton number $Mo=5.4 \times 10^{-7}$, close to that of silicone oil DMS-T05. According to figure \ref{fig:traj_sum}$(a)$, the associated conditions are by far supercritical, since the critical Bond number for this specific liquid is close to $0.85$. 

\begin{figure}
\centerline{\includegraphics[scale=0.65]{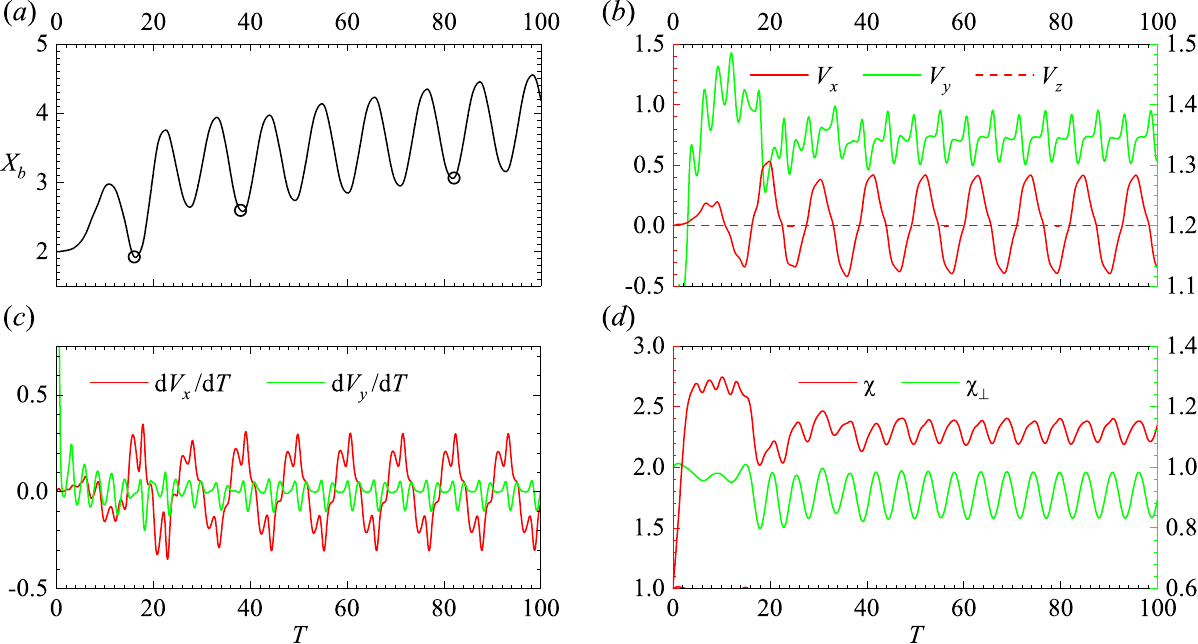}}
\vspace{-1mm}
\caption{Evolution of various characteristics of the bubble path and geometry during the lateral migration of a bubble with $(Bo, Ga) = (1.5, 50)$. ($a$): wall-normal position of the centroid; ($b$): wall-normal (solid red, left axis), horizontal wall-parallel (dashed red, left axis), and vertical (solid green, right axis) components of the bubble centroid velocity; ($c$): wall-normal (red) and vertical (green) acceleration of the centroid; ($d$): principal bubble aspect ratio (red, left axis) and equatorial axes ratio (green, right axis). The three circles in $(a)$ identify the first three moments at which the wake structure is shown in figure \ref{fig:uma_vor1}.
}
\label{fig:uma_motion1}
\end{figure}
\begin{figure}
\centerline{\includegraphics[scale=1.0]{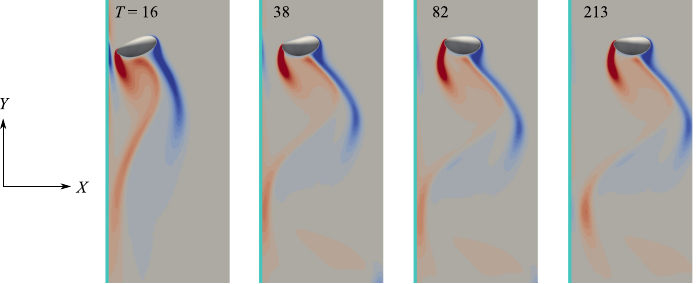}}
\caption{Distribution of the spanwise vorticity, $\overline{\omega}_z$, in the symmetry plane $Z = Z_b$ past a bubble with $(Bo, Ga) = (1.5, 50)$ at several moments when it reaches the wall-facing extremity of a zigzag (circles in figure \ref{fig:uma_motion1}$(a)$). 
Red and blue iso-contours refer to positive and negative values of $\overline{\omega}_z$, respectively, with a maximum magnitude of $2.0$.
}
\label{fig:uma_vor1}
\end{figure}

Figure \ref{fig:uma_motion1} shows how the characteristics of the path and geometry of this bubble evolve as it rises. Panels $(a-b)$ indicate that path instability quickly sets in and saturates after a few cycles of oscillations. Path oscillations take place in the wall-normal plane and give rise to a planar zigzagging motion, since the horizontal wall-parallel velocity component, $V_z$, remains vanishingly small throughout the bubble ascent. At the same time, the bubble gradually migrates away from the wall with, according to panel $(a)$, an average drift velocity close to $0.01$, much smaller than the maximum of $V_x$, which is close to $0.4$. The evolution of this velocity component is far from being perfectly sinusoidal, indicating that the dynamics of the zigzagging motion are nonlinear. This nonlinearity is further highlighted in panel $(c)$, which shows the evolution of the transverse and vertical accelerations. In particular, the time record of $dV_x/dT$ exhibits small oscillations with a frequency five times larger than that of the primary oscillations. The vertical velocity also exhibits strongly non-sinusoidal oscillations which make the bubble rise speed vary upon time by $7.5\%$. 
According to panel $(b)$, the primary oscillations of $V_x$ have a frequency $\overline{f} = 0.092$, in perfect agreement with the predictions of the linear stability analysis performed by \cite{2024_Bonnefis} with a deformable bubble rising in an unbounded fluid (see their figure 10$(b)$). The same study revealed that a secondary mode becomes unstable when $Bo$ exceeds the critical value $1.38$ (their figure 10$(a)$). This mode corresponds to axisymmetric shape oscillations about the bubble minor axis (so-called $(2, 0)$ oscillatory mode for oblate spheroids). The predictions of \cite{2024_Bonnefis} indicate that, for $(Bo, Ga) = (1.5, 50)$, the frequency of these shape oscillations is $5.5$ larger than that of path oscillations, close to the ratio of five noticed in panel $(c)$. This proximity strongly suggests that the high-frequency oscillations present in the evolutions of the velocity and acceleration components in figure \ref{fig:uma_motion1} are the footprint of this mode of shape oscillations.\\
\indent Besides these small-amplitude high-frequency oscillations, figure \ref{fig:uma_motion1}$(d)$ makes it clear that the bubble undergoes significant periodic shape variations along its path. Both the principal aspect ratio, $\upchi$, and the equatorial axes ratio, $\upchi_\perp$, reach their minimum at the extremities of each zigzag, where $V_x$ vanishes and $dV_x/dT$ reaches its extrema. These simultaneous variations indicate that the bubble experiences a periodic compression/dilatation along the equatorial axis lying in the wall-normal plane (the $x'$-axis with length $b$ in figure \ref{fig:problem-state}$(b)$). Examination of bubble contours in the appropriate planes (not shown) reveals that these variations are accompanied by simultaneous phase-opposed oscillations along the $z'$-equatorial axis, while the bubble keeps an almost constant length along the $y'$-minor axis. Therefore, the observed changes in $\upchi$ and $\upchi_\perp$ appear to be driven by pressure variations whose wavelength is half the perimeter of the bubble contour lying in the equatorial $(x',z')$-plane. These oscillations are induced by (and enslaved to) the transverse motion of the bubble. So, their dynamics are distinct from those of natural shape oscillations of bubbles rising in a straight line, as underlined by their much smaller frequency. Note that the wall plays no role in these dynamics, since the figure makes it clear that the minima reached by $\upchi$ and $\upchi_\perp$ are identical at both ends of a given zigzag, i.e. they do not depend on the instantaneous wall-normal position of the bubble. In addition to these zigzag-driven oscillations, the bubble shape exhibits a marked asymmetry along the $x'$-axis, with its vertical cross section taking an egg-like shape pointing towards the exterior of the zigzag (see figure \ref{fig:uma_vor1}). As analyzed by \cite{2016_Cano-Lozano}, this asymmetry results from the periodic rotation of the bubble about the $z'$-axis, which lowers (increases) the pressure on the part of the surface located the farthest from (the closest to) the zigzag centerline.\\ 
\begin{figure}
\centerline{\includegraphics[scale=0.65]{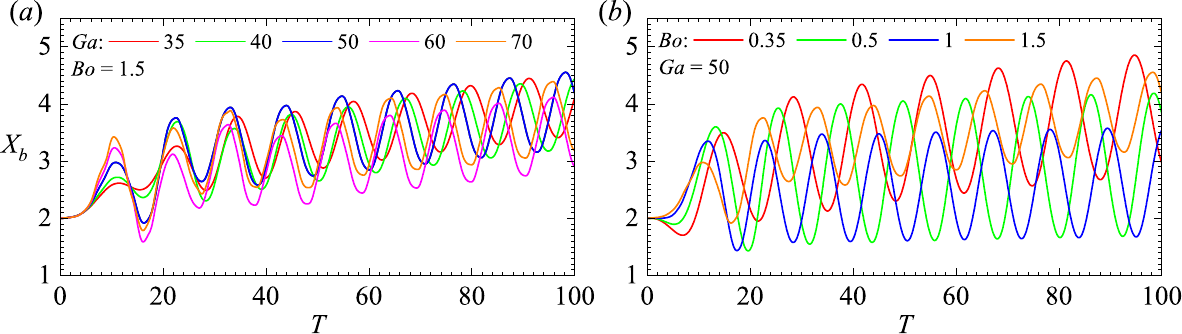}}
\vspace{-1mm}
\caption{Influence of the Bond and Galilei numbers on the evolution of the wall-normal position of the bubble centroid in the WMA regime. $(a)$ increasing $Ga$ at fixed $Bo$; $(b)$ increasing $Bo$ at fixed $Ga$.
}
\vspace{-22.5mm}
\vspace{22.5mm}
\label{fig:uma_mig}
\end{figure}
\indent Figure \ref{fig:uma_vor1} shows how the distribution of the spanwise vorticity evolves as the bubble moves progressively away from the wall. The four snapshots are taken at moments when the bubble reaches the wall-facing extremity of a zigzag. A wall layer with significant $\overline{\omega}_z$-values is present, even ahead of the bubble, at $T=16$, with in particular a stripe of intense negative vorticity (corresponding to a downstream flow) in the bubble-wall gap. This structure still subsists at $T=38$, albeit with a much weaker intensity, and disappears in later times. Only in the downstream region located several radii downstream of the bubble do significant nonzero values of $\overline{\omega}_z$ subsist close to the wall. This is where the wake still weakly interacts with the wall, generating a small repulsive force responsible for the average migration through the familiar vortical mechanism summarized in \S\,\ref{sec:intro} and already active in finite-to-moderately inertial regimes. To summarize, apart from the early stages of the rise, present observations only detect weak interactions of the wake with the wall, and reveal a close agreement with the global stability predictions of \cite{2024_Bonnefis} in an unbounded fluid regarding the characteristics of the zigzagging motion. Combining both aspects leads us to the conclusion that the WMA regime consists essentially of the superposition of the oscillating path resulting from the path instability mechanism in an unbounded flow (with the wall only dictating the orientation of the symmetry plane of the path), and the gradual wake-induced migration away from the wall already active at lower $Ga$, especially with moderately deformed bubbles (see figure \ref{fig:phase-diagram_glob} below). The same conclusion holds in the WMA scenario observed with spiralling bubbles. In particular, we examined the path and dynamics of a bubble with $(Bo, Ga) = (2, 60)$ rising in silicone oil DMS-T05 and found that once its wake, made of a pair of vortex threads wrapped around one another, is fully developed, it only weakly interacts with the wall in a manner similar to that displayed in figure \ref{fig:uma_vor1}.\\
\indent Figure \ref{fig:uma_mig} gathers some records of the bubble path for various $Ga$ at $Bo=1.5$ (panel $(a)$), and for various $Bo$ at $Ga=50$ (panel $(b)$). Clearly, the average velocity with which the bubble migrates away from the wall varies non-monotonically with the control parameters in both cases. For instance, the fastest migration in panel $(a)$ is seen to be reached for $Ga=50$ (blue curve), while the slowest one is obtained for $Ga=60$ (purple), with the two evolutions corresponding to $Ga=40$ (green) and $70$ (orange) exhibiting intermediate migration velocities. Noting that in the saturated stage the amplitude of the zigzagging motions experience little variation with $Ga$, it appears that the early stages of the path are responsible for the most part of the non-monotonic variations of the average migration velocity. For a fixed $Bo$, the higher $Ga$ the larger the growth rate of the zigzagging motion, since increasing $Ga$ increases the distance to the path instability threshold. For this reason, the transverse oscillations of paths corresponding to $Ga=60$ and $70$ quickly reach a large amplitude, which, at the end of the first cycle of oscillations, brings the bubbles closer to the wall than at the time of their injection (e.g. $X_b\approx1.6$ at $T=16$ for $Ga=60$). Taking the bubble with $Ga=35$ as reference, this induces a delay in the average migration of bubbles with a higher $Ga$ that is never caught up in later stages. Similarly, when $Ga$ is kept fixed and $Bo$ is increased, the bubble closest to the path instability threshold ($Bo=0.35$, red line in panel $(b)$) quickly reaches average distances from the wall larger than bubbles with $Bo=0.5$ (green) and $Bo=1$ (blue), whose zigzags amplitude grows faster. However, at saturation, these amplitudes greatly vary with the Bond number, i.e. with the bubble shape, passing through a maximum for $Bo=0.5$ (see \S\,\ref{sec:nwz_mec} for more discussion on this aspect). As a result of this marked variation, the amplitude of the zigzags performed by the bubble with $Bo=0.5$ is twice as large as that of the bubble with $Bo=1.5$. This is why the latter migrates faster than those with $Bo=1$ and $0.5$ which spend more time close to the wall throughout their ascent. 

\subsection{NWZ regime}
\label{sec:nwz}
\begin{figure}
\centerline{\includegraphics[scale=0.65]{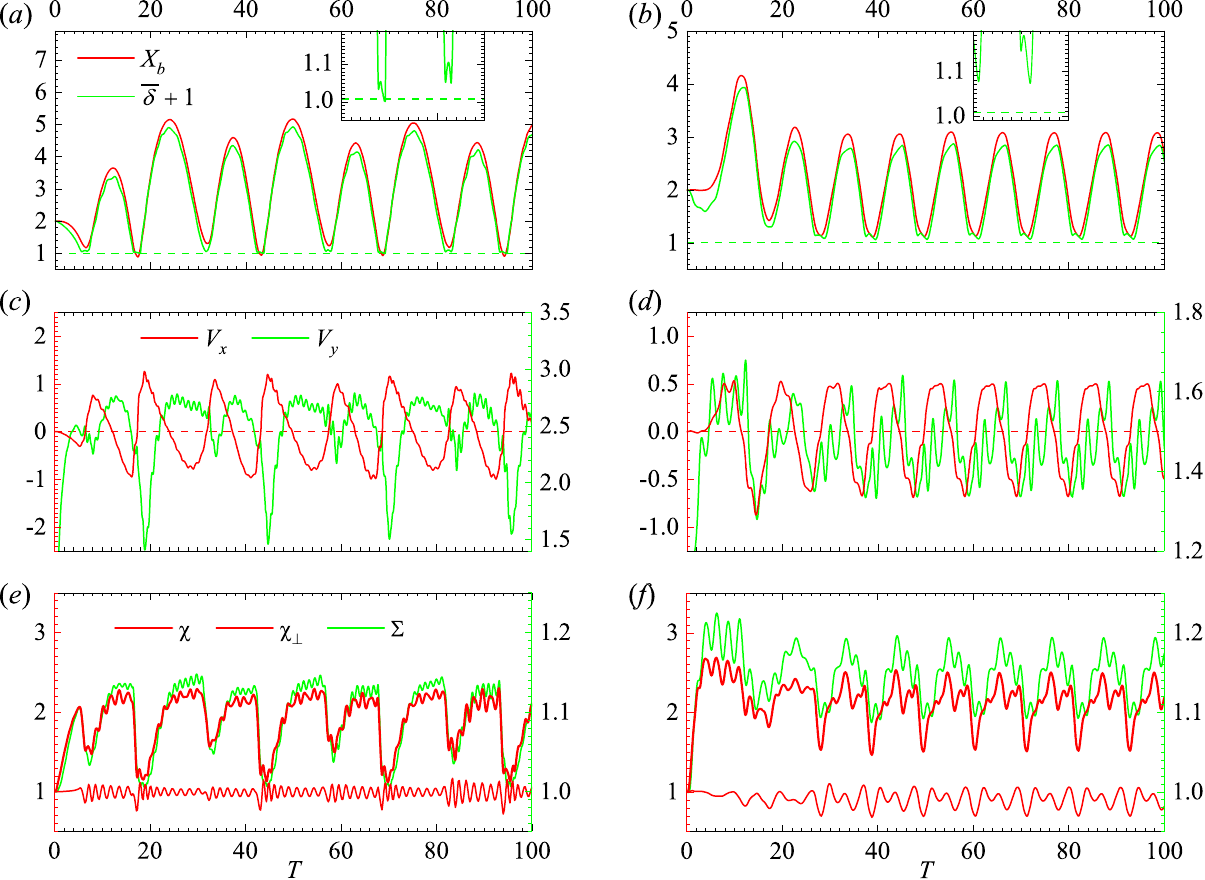}}
\vspace{-1mm}
\caption{Evolution of some characteristics of the bubble dynamics for $(Bo, Ga) = (0.25, 90)$ (left column) and $(1, 70)$ (right column). $(a,b)$: wall-normal position of the bubble centroid (red line) and gap thickness (green line); $(c,d)$: wall-normal (red) and vertical (green) components of the velocity of the bubble centroid; $(e,f)$: principal aspect ratio (thick red line), equatorial axes ratio (thin red line) and surface area (green line). In $(c-f)$, the right axis refers to the quantity shown with the green line. In $(a-b)$, the insets are located at the actual position on the horizontal (time) axis and only their vertical axis is stretched; the green dashed line identifies the transverse position $1+\overline\Delta_\text{min}$.}
\label{fig:nwz_motion}
\end{figure}

Figure \ref{fig:traj_sum}$(a)$ indicates that, beyond the neutral curve but some distance away from it, bubbles with $Ga\gtrsim60$ and $Bo\lesssim1$ follow a near-wall zigzagging (NWZ) scenario. Figure \ref{fig:nwz_motion} illustrates the evolution of some characteristics of the dynamics of two such bubbles. The main difference between the two evolutions is the occurrence of direct bubble-wall collisions in the former case. Collision events can be identified from the temporal evolution of the dimensionless gap, $\overline\delta$. In line with the definition recalled in \S\,\ref{sec:problem_state}, we consider that a direct collision takes place when $\overline\delta$ becomes less than the minimum cell size, $\overline\Delta_\text{min}$, such as during the first event displayed in the inset of figure \ref{fig:nwz_motion}$(a)$. As this figure shows, $\overline\delta$ decreases to $\overline\Delta_\text{min}$ at regular time intervals, first at $T \approx 18$. In all NWZ evolutions involving collisions, these events occur every other round of transverse oscillations. Upon collision, $V_x$ reverses and peaks at an absolute value larger than that achieved prior to the collision (see panel $(c)$), which may be interpreted as a rebound with a restitution coefficient larger than one. Consequently, the maximum separation reached by the bubble in the upcoming zigzag is significantly larger than that achieved just before the collision. In contrast, when collisions do not take place, the maximum separation remains constant once the oscillations have saturated (see panel $(b)$), and the restitution coefficient is smaller than one throughout the bubble ascent. Collisions also deeply affect the evolution of the bubble rise speed. As panel $(c)$ shows, this component reduces by nearly $50\%$ shortly after the collision ($V_y\approx1.5$). In contrast, this reduction reaches only $20\%$ in the next zigzagging cycle during which no collision takes place, and the same observation holds for the second bubble that never collides with the wall (panel $(d)$). In addition to this dominant variations enslaved to the zigzag dynamics, $V_y$ also exhibits fast small-amplitude oscillations, especially in the case of the bubble with the smallest $Bo$. Qualitatively similar evolutions are noticed in panels $(e-f)$ for the principal aspect ratio $\upchi$, and the surface area $\Sigma$ (defined as the area of the gas-liquid interface normalized by $4\pi R^2$). These results reveal the existence of two distinct modes of shape oscillations: a primary low-frequency mode, by which $\upchi$ and $\Sigma$ reach their maximum (respectively, minimum) when the bubble-wall separation achieves its maximum (respectively, minimum), and a secondary high-frequency mode with a significantly smaller amplitude. This second mode is seen to dominate the time variations of the equatorial axes ratio $\upchi_\perp$, whose relative magnitude is up to $20\%$ in the stages when the bubble is `far' from the wall. 
We compared the frequency of this second mode with that of small-amplitude capillary oscillations in the inviscid limit. For a nearly-spherical bubble, the fundamental mode of such oscillations has a frequency $\overline{f}_2 = \sqrt{12}Bo^{-1/2}$ \citep{1932_Lamb}. Increasing the bubble oblateness leads to a decrease in this frequency, as the nonlinear computations of \citet{1989_Meiron} showed. The two deformable bubbles considered in figure \ref{fig:nwz_motion} have a principal aspect ratio $\upchi \approx 2.1$. For this oblateness, Meiron's predictions 
indicate that the frequency of the lowest-order axisymmetric oscillations (corresponding to the $(2,0)$ mode) is $\approx0.84\overline{f}_2$. However, variations observed in panels $(e-f)$ on the equatorial axes ratio $\upchi_\perp$ prove that the observed high-frequency oscillations are three-dimensional. The lowest two three-dimensional capillary modes are the $(2,1)$ and $(2,2)$ ones, the wavelength of which in the equatorial plane is the corresponding perimeter and half of it, respectively. 
Meiron's results predict that, for an oblate spheroid with $\upchi=2.1$,  their frequency is  approximately $0.81\overline{f}_2$ and $0.56\overline{f}_2$, respectively. Inspection of the records of $\upchi_\perp$ in panels $(e-f)$ indicates that the frequency of the observed high-frequency oscillations is close to $0.59\overline{f}_2$ at $Bo=0.25$ and $0.50\overline{f}_2$ at $Bo=1$. Given that the rise Reynolds number in the second case is less than half that in the first one, viscous corrections are expected to be more significant in the former. With this in mind, this comparison strongly suggests that the observed oscillations correspond to the non-axisymmetric $(2,2)$ mode for both bubbles.

\begin{figure}
\centerline{\includegraphics[scale=0.65]{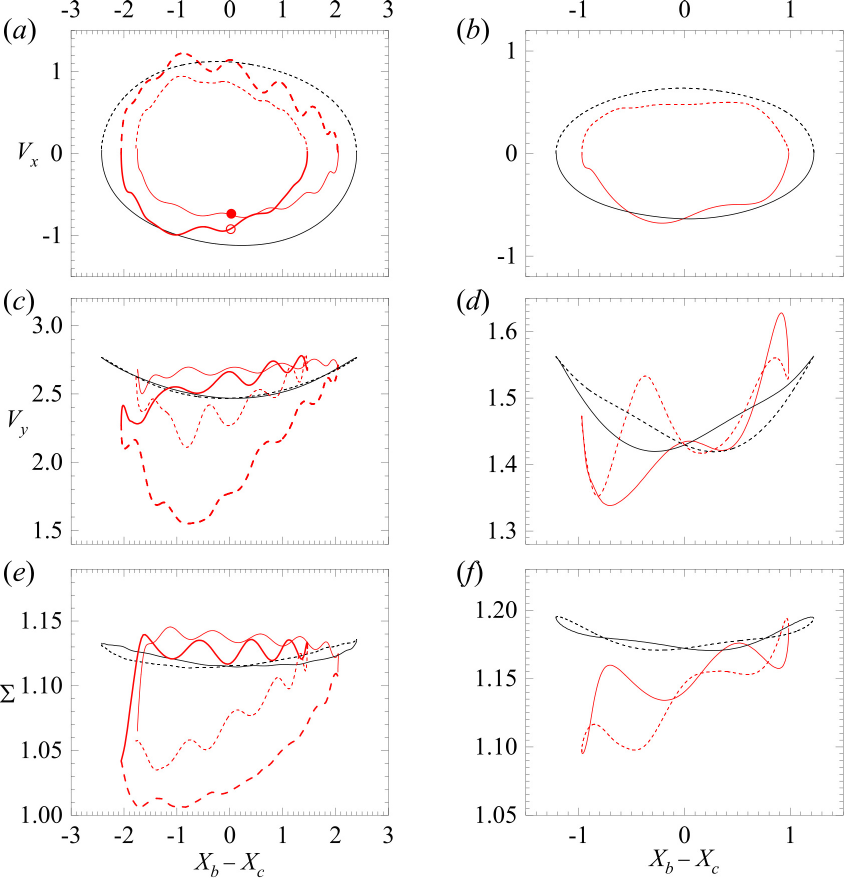}}
\vspace{-1mm}
\caption{Variation of several characteristics of the bubble dynamics over a single zigzag period in the presence or absence of a wall for two bubbles with $(Bo, Ga) = (0.25, 90)$ (left column) and $(Bo, Ga) = (1, 70)$ (right column). $(a,b)$: horizontal velocity component ,$V_x$, lying in the plane of the zigzagging motion (i.e., wall-normal component when the wall is present); $(c,d)$: vertical velocity, $V_y$; $(e,f)$: surface area, $\Sigma$. Red and black lines denote results in the presence and the absence of the wall, respectively. Solid and dashed lines refer to the half-period of the zigzag with negative and positive $V_x$, respectively; thick and thin red lines correspond to the sub-period of the zigzag with and without a collision, respectively.}
\label{fig:nwz_wb_vs_ub}
\end{figure}
To better quantify wall effects, we carried out additional runs considering the same two bubbles in an unbounded domain. After exhibiting flattened spiralling paths, they both eventually describe planar zigzags with reduced frequencies $St=0.057$ and $St=0.113$ for $(Bo, Ga) = (0.25, 90)$ and $(1, 70)$, respectively. These frequencies are slightly smaller than their wall-bounded counterparts ($St=0.063$ and $0.126$, respectively). However, the most prominent difference is that shape oscillations keep a much smaller amplitude in the absence of the wall. This difference is highlighted in figure \ref{fig:nwz_wb_vs_ub} which shows the variation over a single zigzag period of the velocity components of the bubble centroid and its surface area as a function of the lateral displacement, $X_b - X_c$, with $X_c$ being the time-averaged wall-normal position of the centroid, i.e. the horizontal position of the zigzag centreline. In the unbounded configuration, the two halves of the zigzag period exhibit symmetric evolutions with respect to the mean position $X = X_c$. For both bubbles, the relative change in $\Sigma$ remains less than 3\% throughout the zigzag period, indicating weak shape oscillations. 
In the presence of the wall, the mean amplitude of the lateral oscillations is reduced by $20$ to $25\%$ and no symmetry with respect to $X=X_c$ subsists. The bubble surface area experiences marked variations, with relative changes from $10$ to $15\%$. For both bubbles, $\Sigma$ attains its minimum shortly after the bounce, making the bubble less deformed during most of the departing stage than when it recedes to the wall.  

\begin{figure}
\centerline{\includegraphics[scale=0.65]{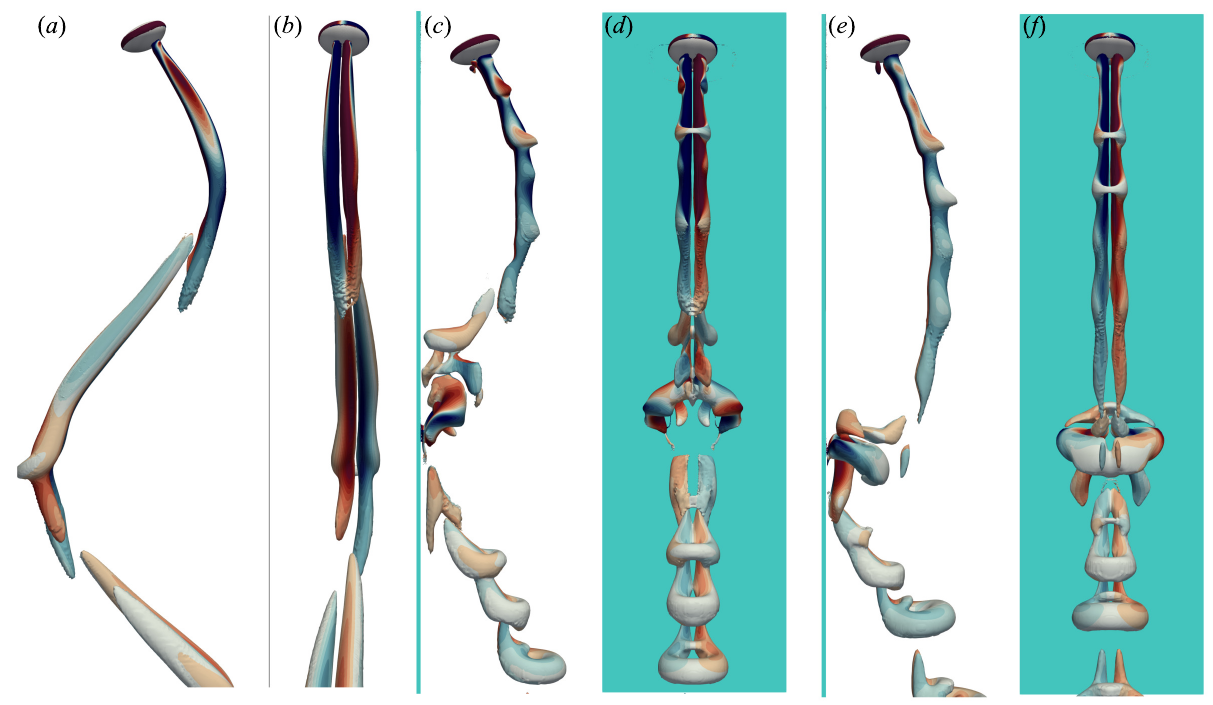}}
\caption{Wake structure past a zigzagging bubble with $(Bo, Ga) = (0.25, 90)$ at the moment when the bubble crosses the centreline of the zigzag. $(a-b)$: unbounded configuration; $(c-f)$: wall-bounded configuration. Snapshots in panels $(c-d)$ and $(e-f)$ are taken at the instants of time marked with open and closed circles in figure \ref{fig:nwz_wb_vs_ub}$(a)$, respectively. The wake structure is visualized using the $\lambda_2$ criterion, with iso-$\lambda_2$ surfaces colored by the local value of $\overline{\omega}_y$; red and blue contours indicate positive and negative values, respectively, with a maximum magnitude of $1.0$. The wall is indicated by a dark green line in $(c,e)$ and a dark green rectangular surface in $(d,f)$.}
\label{fig:nwz_wake}
\end{figure}
Figure \ref{fig:nwz_wake} visualizes the wake structure downstream of the bubble with $(Bo, Ga) = (0.25, 90)$ using the $\lambda_2$ criterion \citep{1995_Jeong}. 
In the unbounded case (panels $(a,b)$), this structure corresponds to the `4R' vortex shedding mode, following the terminology of \citet{2010_Horowitz}. This mode refers to two pairs of counter-rotating vortices with opposite signs of $\overline{\omega}_y$ successively shed in the wake during a zigzag half-period, with one pair being much stronger than the other, thus allowing the path to remain almost sinusoidal. 
The `4R' mode is known to be linked to oscillations in the relative velocity between the body and fluid (here the bubble rise speed). It emerges when the relative magnitude of these oscillations (with reduced frequency $2St$) exceeds approximately $10\%$ \citep{2010_Horowitz, 2016_Cano-Lozano, 2018_Auguste}, a condition widely fulfilled in the present case according to figure \ref{fig:nwz_motion}$(c)$. 
In the wall-bounded configuration (panels $(c-f)$ in figure \ref{fig:nwz_wake}), the vortex pair shed towards the wall degenerates into a short series of vortex patches, creating a strong suction effect on the bubble in the early departure stage. This is the cause of the sharp decrease in the bubble rise speed and surface area during this stage. The wall effect is less severe in cases where collisions do not happen, such as the bubble with $(Bo,Ga)=(1,70)$ (not shown). In such cases, the wall compels the same vortex pair to bend and align vertically but the corresponding primary vortices do not break into distinct patches. \\
\indent The structure of the vortex pair shed towards the fluid interior differs significantly from that in the unbounded configuration. Specifically, several single-sided, short-wavelength loops superimpose onto the primary vortex pair (panels $(c,e)$ in figure \ref{fig:nwz_wake}). These loops grow in time and progressively invade the far-wake structure. Close inspection indicates that their number matches that of the high-frequency oscillations of the bubble aspect ratio, suggesting that these short-wavelength loops are a by-product of shape oscillations. Indeed, every transient change in the bubble shape (especially in the vicinity of the bubble's equator) results in a change in the local curvature of the bubble surface, hence in a variation of the magnitude of the tangential surface vorticity \citep{2008_Veldhuis, 2016_Cano-Lozano}. The lifetime of these short-wavelength loops depends on the oscillatory Reynolds number, $\Rey_{osc}=(\rho_l\gamma R)^{1/2}/\mu_l=Ga\,Bo^{-1/2}$, that compares the characteristic time of shape oscillations to that of their viscous damping. Hence, when the Bond number is low enough, these oscillations decay slowly and are able to alter significantly the far-wake structure. 

\subsection{How a zigzagging bubble gets trapped near the wall}
\label{sec:nwz_mec}
That the NWZ regime is encountered in an intermediate range of Bond numbers for $Ga\geq60$ is somewhat surprising at first glance. Indeed, at a given $Ga$, bubbles with a slightly subcritical $Bo$ migrate away from the wall and those whose Bond number exceeds $1.5$ do the same  through the WMA scenario (figure \ref{fig:traj_sum}$(a)$). The reduced deformation of the bubble in the departing stage of the zigzagging motion revealed by figures \ref{fig:nwz_wb_vs_ub}$(e,f)$, which itself results from the reduction in the rise speed induced by the intense interaction of the wake vortices with the wall (figures \ref{fig:nwz_wake}$(c-f)$), is responsible for the observed trapping phenomenon. Indeed, since the minor and major axes of an oblate spheroid vary as $\upchi^{-2/3}$ and $\upchi^{1/3}$, respectively, the frontal area of the bubble involved in the transverse motion, say $S_\perp$, varies as $\upchi^{-1/3}$. Thus, it is larger when the bubble moves away from the wall with $\upchi\approx1$ than when it goes back to it with $\upchi>2$.
Throughout the departing stage, this results in an increase in the transverse viscous drag (for a given wall-normal velocity), as well as in the amount of liquid displaced transversely by the bubble, hence the transverse added-mass force (for a given wall-normal acceleration). 
Both effects tend to oppose the departing motion of the bubble. Therefore, the larger the frontal area $S_\perp$, the weaker the positive transverse velocity $V_x$ the bubble can achieve at a given wall-normal position $X_b$. This is why, as figure \ref{fig:nwz_wb_vs_ub} highlights, the amplitude of the zigzags, i.e. the horizontal distance between two successive $V_x$-reversals, is reduced compared to that achieved by the same bubble in an unbounded fluid. Moreover, in the highly-inertial regime under consideration, the repulsive interaction force is known to vary approximately as $X_b^{-4}V_y^2$ for a nearly spherical bubble \citep{2003_Takemura}. Although bubbles concerned by the NWZ regime are far from spherical, this scaling provides a strong indication that the dramatic reduction in the rise speed during the first half of the departing stage leads to a drastic drop in the repulsive force. Both aspects cooperate to hamper the drift of the wall-normal position of the zigzag centreline, $X_c$, yielding a near-wall trapping of the bubble from which the observed periodic path ensues. 

\begin{figure}
\centerline{\includegraphics[scale=0.65]{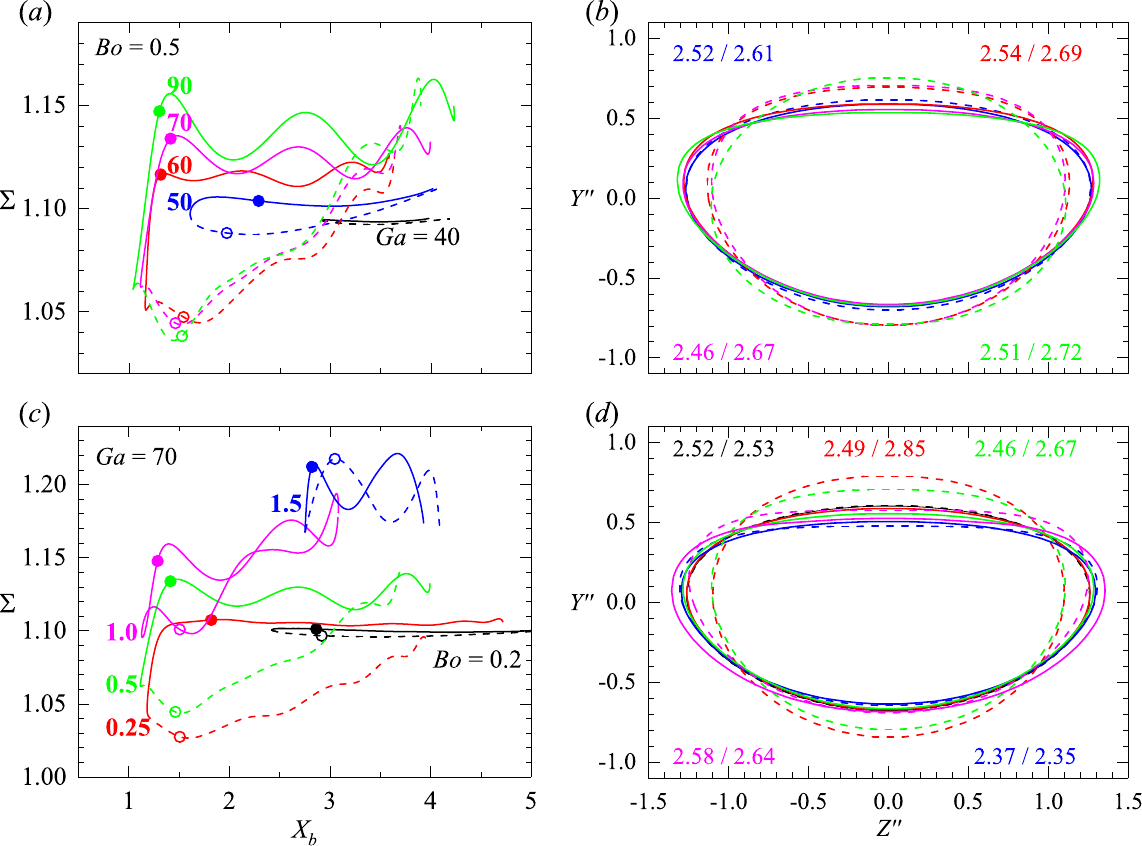}}
\vspace{-1mm}
\caption{Time-dependent deformation of bubbles obeying a NWZ or a WMA scenario. $(a)$: surface area, $\Sigma$, vs the wall distance, $X_b$, over one pseudo-period of the zigzag for $Bo=0.5$ and various $Ga$; $(b)$: 
bubble cross section in the diametrical plane $(Y",Z")$ perpendicular to the transverse motion at $X_b$ positions midway between the centreline of the zigzag and its extremity closest to the wall (these positions are marked with circular symbols in panel $(a)$); the $Y"$ axis is parallel to the bubble minor axis and the $Z"$ axis is horizontal and parallel to the wall. $(c,d)$: same as $(a,b)$ for $Ga=70$ and various $Bo$. In all panels, the black and blue evolutions belong to the WMA regime. 
Solid (respectively, dashed) lines refer to the approaching (respectively, departing) stage. Values shown in $(b,d)$ correspond to the frontal area,  $S_\perp$, enclosed in the contour of the same colour, with the first and second numbers in each pair referring to the approaching and departing stages, respectively. 
}
\label{fig:nwz_mec2}
\end{figure}
Figure \ref{fig:nwz_mec2}$(a)$ illustrates the variations of the bubble area, $\Sigma$, with the wall distance over one zigzag pseudo-period for bubbles with $Bo = 0.5$ at different $Ga$. With no surprise, $\Sigma$ experiences negligible changes at $Ga = 40$, the critical $Ga$ at which path instability sets in at this specific Bond number. These variations become discernible at $Ga = 50$ but the bubble still manages to escape from the wall region. The surface area is seen to experience much larger changes (of the order of $7-8\%$) for $Ga\geq60$. Figure \ref{fig:traj_sum}$(a)$ indicates that this corresponds to the $Ga$-range in which the NWZ regime is observed for $Bo = 0.5$, which is in line with the physical arguments presented above. Figure \ref{fig:nwz_mec2}$(b)$ shows how the contour of the bubble in the diametrical plane perpendicular to the transverse motion varies along the departing and approaching stages at a given $Ga$. These plots confirm that this cross section is significantly less oblate in the departing stage for $Ga>50$. The relative difference in the corresponding frontal area between the two stages becomes more pronounced as $Ga$ increases further. This difference increases with $Ga$, i.e. with the distance to the threshold of path instability, and is about $8\%$ at $Ga = 90$. 
Figures \ref{fig:nwz_mec2}$(c,d)$ show results similar to those in the previous two panels for bubbles with $Ga = 70$ over a wide range of Bond number. Almost no change in either $\Sigma$ or $S_\perp$ is noticed at $Bo = 0.2$, the critical $Bo$ at which path instability sets in at this specific Galilei number. The situation changes drastically when the Bond number increases up to $0.25$, with now a relative difference of around $15\%$ in $S_\perp$ between the approaching and departing stages. Then, this difference reduces gradually as $Bo$ goes on increasing. The NWZ regime subsists up to $Bo=1$ and is succeeded by the WMA regime at $Bo=1.5$, where $\Sigma$ and $S_\perp$ both experience negligible variations. 

As pointed out above, the NWZ regime is a consequence of the intense interaction of the wall with the double-threaded wake accompanying zigzagging bubbles. Therefore, for this regime to exist, it is necessary that the lateral excursions performed by the zigzagging bubble have a sufficient amplitude. Given the supercritical nature of path instability \citep{2002b_Mougin}, the required amplitude may only be reached if $Bo$ is somewhat larger than $Bo_c(Ga)$, the critical Bond number corresponding to the onset of the zigzagging motion at the considered $Ga$. This is why bubbles with $Bo$ only slightly above $Bo_c$ are still able to migrate away from the wall. The reason for the existence of a maximum $Bo$ for the NWZ regime is somewhat more subtle. As is well known, zigzagging bubbles rise in such a way that their minor axis remains almost aligned with their path at all times \citep{2001_Ellingsen, 2006_Mougin}. For this, they perform oscillatory rigid-body rotations with an angular velocity that vanishes at the inflection points of the zigzags and reaches its maxima at their extremities. The more oblate the bubble is, hence the larger $Bo$, the larger its resistance to such rotational motions is. This is due on the one hand to the rapid increase of the viscous torque with $\upchi$ (for a given rotation rate), and on the other hand to that of the moment of inertia of the liquid entrained by the rotational motion (so-called rotational added-mass coefficient), which governs the rate of change of the angular velocity. For instance, both quantities almost double from $\upchi=2$ to $\upchi=2.5$ when $\Rey\gg1$ \citep{2011_Magnaudet}. This sharp increase limits severely the amplitude of the oscillatory rotations a bubble with a given volume may perform, which translates directly into a reduction in the amplitude of the zigzags. For instance, with $Ga=70$, present simulations indicate that this amplitude decreases by a factor of two from $Bo=0.25$ to $Bo=1.5$ in the unbounded configuration, and a similar decrease may be observed in figure \ref{fig:nwz_mec2}$(c)$ in the presence of the wall. The above arguments concur to the conclusion that the NWZ regime can only exist within a finite range of $\upchi$, hence of $Bo$. 

\section{Summary}
\label{sec:conclusion}

We carried out three-dimensional numerical simulations of the buoyancy-driven motion of freely deformable bubbles rising near a vertical wall in the parameter range $0.02 \leq Bo \leq 2, 35 \leq Ga \leq 90$. Within this range, provided the Bond number exceeds a $Ga$-dependent threshold, $Bo_c(Ga)$, an isolated bubble immersed in an unbounded fluid follows a zigzagging or spiralling path, whereas it rises in a straight line if $Bo<Bo_c(Ga)$.
In the latter case, the three distinct regimes of near-wall rising motions discussed at smaller $Ga$ in Part 1, namely periodic near-wall bouncing, damped bouncing, and migration away from the wall, are observed up to $Ga \approx 50$.\\ 
\indent For larger $Ga$ and small Bond numbers (typically $Bo<0.05$ at $Ga = 60$ and $Bo\lesssim0.1$ for $Ga \geq 70$), a new regime is found. In that bouncing-tumbling-escaping (BTE) scenario, the bubble manages to escape from the wall region after one to two rounds of near-wall bouncing. The escape mechanism is rooted in the abrupt flow reversal that takes place in the gap just before the bubble collides with the wall. This reversal is a consequence of the drop in the bubble rise speed which brings part of the fluid previously contained in the wake into the gap region. As a result, a strong rotational flow forms around the bubble surface, leading to a repulsive Magnus-like lift force that keeps a magnitude of the same order as the buoyancy force over a significant period of time after the collision. Bubbles are found to eventually escape from the wall region every time the spinning rate characterizing the rotational flow at the bubble surface exceeds a critical value. A crude force balance allows the coefficient involved in the expression of the Magnus-like force to be estimated at around $0.5$.\\
\indent In an unbounded domain, path instability takes place when $Bo \geq Bo_c(Ga)$. Paths of bubbles rising near a vertical wall are also found to experience transverse oscillations as soon as $Bo > Bo_c(Ga)$. This suggests that the wall plays little role in the occurrence of path instability. Its most significant effect appears to be the selection of the plane where the zigzagging motion takes place, which is always perpendicular to the wall. Simulations reveal that zigzagging or spiralling bubbles rising near a wall experience one of the following two scenarios.\\
\indent For $Bo > Bo_c(Ga)$ and $Ga$ up to $50$, or $Bo > 1$ and larger $Ga$, bubbles ultimately migrate away from the wall, following a zigzagging motion for $Bo \lesssim 1.5$ and a spiralling motion at larger $Bo$, on which a small lateral drift superimposes. The same phenomenology is observed whatever $Ga$ just above the critical curve, i.e. for $Bo\gtrsim Bo_c(Ga)$. In this wavy migration away (WMA) regime, the characteristics of the transverse path oscillations are very close to those determined in an unbounded domain. This is especially true regarding frequencies, be they those of the path oscillations themselves or those of the shape oscillations that develop under sufficiently supercritical conditions. In this regime, bubble wakes only interact weakly with the wall as soon as the gap thickness exceeds half of the bubble radius. Therefore, the repulsive interaction mechanism is essentially similar to that observed below the path instability threshold at lower $Ga$. The net migration away from the wall does not vary monotonically with the control parameters $Ga$ and $Bo$. This is mostly due to the initial stages of the rise, during which the amount of time a given bubble spends close to the wall depends dramatically on the growth rate and saturated amplitude of the zigzags. For this reason, bubbles following paths with a slowly growing zigzagging component or performing lateral excursions with a moderate amplitude migrate faster.\\ 
\indent For $Ga \geq 60$ and intermediate Bond numbers slightly larger than $Bo_c(Ga)$ but smaller than an upper value between $1$ and $1.5$, bubbles maintain a near-wall zigzagging (NWZ) motion without migrating towards the bulk. The trapping phenomenon characterizing this regime is a consequence of the intense interaction of the double-threaded wake with the wall during the stages when the gap comes to its minimum. The energy dissipation resulting from this wake-wall interaction translates into a severe drop in the rise speed, which in turn results in a drastic transient reduction of the bubble oblateness. Because of this, the frontal area opposing the transverse motion is larger when the bubble departs from the wall than when it returns to it. This makes it possible for resistive transverse forces, such as viscous drag and added mass, to counteract the repulsive interaction force, maintaining the centreline of the zigzagging motion a constant distance from the wall. The  mechanism at play requires the zigzag amplitude to be large enough for bubbles to be trapped. This is why this regime only exists up to a maximum Bond number, the lateral excursions of highly oblate bubbles being severely limited by viscous and inertial effects resisting  the periodic rotation needed to keep such bubbles broadside on along their path.  

\begin{figure}
\centerline{\includegraphics[scale=0.67]{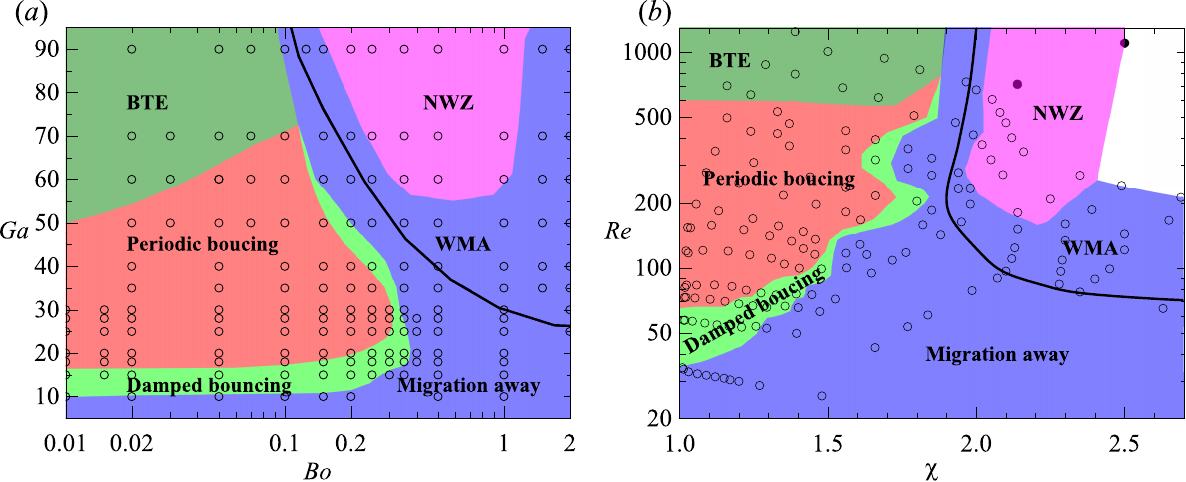}}
\caption{Complete state diagram of near-wall rising regimes observed in the simulations up to $Ga=90$ and $Bo=2$. $(a)$: $(Bo,Ga)$-plane; $(b)$: $(\upchi,\Rey)$-plane, with $\upchi$ and $\Rey$ determined as explained in the caption of figure \ref{fig:chi-vs-re}. Solid line: neutral curve corresponding to the onset of path instability in an unbounded fluid (\citet{2024_Bonnefis}. The blue zone straddling the neutral curve represents the whole set of conditions under which bubbles migrate away from the wall, either in the presence or in the absence of path instability. In $(b)$, the two bullets at $(\upchi, \Rey) = (2.1, 710)$ and $(\upchi, \Rey) = (2.5, 1100)$ correspond to experimental data from \cite{2001_Vries} and \cite{2015_Jeong}, respectively. Both were obtained in water and show the persistence of the NWZ regime beyond the maximum $Ga$ reached in the simulations.}
\label{fig:phase-diagram_glob}
\end{figure}
\indent Figure \ref{fig:phase-diagram_glob} gathers present results and those obtained at lower $Ga$ in Part 1. It provides an overview of all the near-wall rising regimes we could identify for $Ga$ up to $90$ and $Bo$ up to $2$. This map allows a global assessment of the influence of fluid inertia and bubble deformation on the transitions between the various styles of path. For weakly deformed bubbles, say $Bo \lesssim 0.1$, the response of the system is largely controlled by the two antagonistic forces that both originate in fluid inertia, namely the repulsive force induced by the vortical wake-wall interaction mechanism summarized in \S\,\ref{sec:intro} and the attractive force resulting from the acceleration of the fluid in the gap, as predicted when the flow is assumed irrotational. The vortical effect dominates up to $Ga\approx10$, making bubbles migrate away from the wall whatever $Bo$. The influence of the attractive irrotational effect increases gradually with increasing $Ga$ (hence $\Rey$), driving bubbles  towards the wall when $Ga\gtrsim10$ (corresponding to $\Rey \gtrsim 35$). This attractive process gives rise to a damped near-wall bouncing regime up to $Ga\approx17$, i.e. $\Rey\approx65$, then to a periodic near-wall bouncing regime at higher $Ga$. The latter persists up to $Ga \approx 50$ at $Bo = 0.01$ and $Ga \approx 70$ at $Bo = 0.1$, which in both limits corresponds to $\Rey\approx500$. For larger $Ga$, low-$Bo$ bubbles escape from the wall region following the BTE scenario. In this regime, the mechanism responsible for the bubble escape may be viewed as a side effect of the above irrotational mechanism. Indeed, the intensity of the tumbling flow that forms around the bubble surface during the collision with the wall depends crucially on the wall-ward velocity of the bubble and, thus, on the attractive force acting on it.\\
\indent As $Bo$ increases beyond $0.1$, bubbles become more oblate in inertia-dominated regimes. The above two antagonistic effects are still at work, but their magnitude is strongly influenced by the bubble shape. In particular, deformation significantly enhances the repulsive vortical effect, owing to the increase of the tangential vorticity at the bubble surface with the local curvature of this surface. This promotes the migration away from the wall, as the $(\upchi,Re)$ representation in figure \ref{fig:phase-diagram_glob}$(b)$ highlights. Below the path instability threshold, all bubbles with $Ga \leq 30$ (respectively, $Ga>30$) migrate away from the wall as long as their aspect ratio exceeds $\approx1.5$ (respectively $\approx1.8$). In the presence of path instability, i.e. for $\upchi \gtrsim 1.95$, bubbles still exhibit an average migration away from the wall, provided that the transient reduction in $\upchi$ they experience when they get very close to the wall remains weak enough. This reduction becomes significant (say $\geq5\%$) when $Ga \geq 60$ for intermediate Bond numbers such that $Bo_c(Ga)\lesssim Bo\lesssim1$. In this high-$Ga$ intermediate-$Bo$ range, bubbles are trapped near the wall, undergoing a zigzagging-like motion without being able to escape to the bulk.\\
\indent The two parts of this study provide comprehensive insights into the complex dynamics of buoyancy-driven isolated bubbles rising near a vertical hydrophilic wall. This second part highlights the complex interplay between wake-wall interactions, wall-induced fluid displacements, time-dependent bubble deformation and, in the relevant parameter range, lateral bubble excursions resulting from path instability. The findings obtained in this study form a solid basis for developing low-order predictive semi-empirical models of near-wall bubble motion under a broad range of flow conditions. However, to get closer to multiphase flow systems used in engineering applications, it will be particularly relevant to extend the present investigation to configurations involving Marangoni effects. Indeed, temperature differences are frequently imposed along the wall or between the wall and fluid, and, in most systems, bubbles are contaminated by surfactants present in the fluid, especially with water. Thus, how the associated surface tension variations affect the near-wall bubble dynamics over the various flow regimes is a key issue to be considered in a future study.




\backsection[Funding]{P.S. acknowledges the funding of the Deutsche Forschungsgemeinschaft (DFG, German Research Foundation) through grant number 501298479.}

\backsection[Declaration of interests]{The authors report no conflict of interest.}


\backsection[Author ORCIDs]{

Pengyu Shi, https://orcid.org/0000-0001-6402-4720; 

Jie Zhang, https://orcid.org/0000-0002-2412-3617;

Jacques Magnaudet, https://orcid.org/0000-0002-6166-4877.
}


\appendix

\section{Preliminary numerical tests}
\label{sec:app_pre}

Extensive preliminary tests were carried out to assess the reliability of the numerical approach. According to the benchmark test for rising bubbles in \emph{Basilisk} \citep{2017_Popinet}, the temporal and spatial accuracies of the predictions depend largely on five parameters: the Courant-Friedrichs-Lewy (CFL) number $N_{CFL}$, the standard tolerance in the Poisson solver $T_\varepsilon$, the grid refinement criteria for the phase fraction $\zeta_f$ and velocity $\zeta_u$, and the minimum grid size $\overline\Delta_{\min}$. For the specific problem under consideration, the tests conducted in Part 1 confirmed that setting $N_{CFL} = 0.5$, $T_\varepsilon = 10^{-4}$, $\zeta_f = 10^{-3}$, $\zeta_u = 10^{-2}$, and $\overline\Delta_{\min} = 1/68$ (further decreased to 1/136 when $\overline\delta_{\min} \leq 0.15$) provides appropriately converged results. Specifically, the settings for the last two parameters guarantee a sufficient spatial resolution of the boundary layers at the wall and the bubble surface, as well as in the far wake ($\gtrsim10R$ downstream from the bubble centroid) for $Ga$ up to $30$, the highest $Ga$ considered in Part 1. In the present work, $Ga$ goes up to 90, necessitating further verification of the values selected for these two parameters. For this purpose, we consider a case in the BTE regime with $(Bo, Ga) = (0.05, 90)$, for which the bubble Reynolds number (based on the terminal velocity and bubble diameter) is approximately $940$. Therefore, with $\overline\Delta_{\min}=1/68$ (respectively $1/136$) and $\zeta_u = 10^{-2}$, approximately three (respectively six) cells lie in the boundary layer surrounding the bubble surface.

\begin{figure}
\centerline{\includegraphics[scale=0.65]{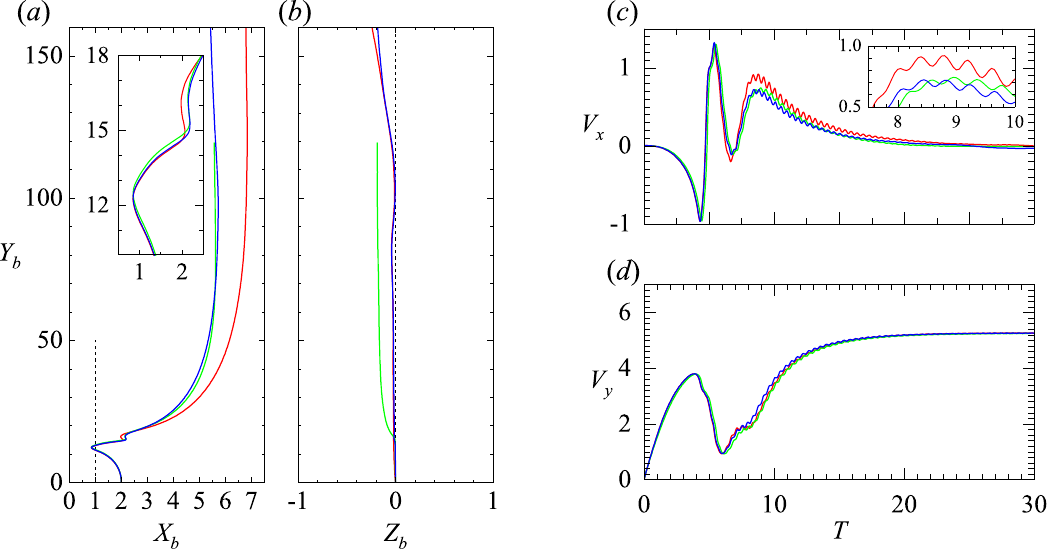}}
\vspace{-1mm}
\caption{Effects of the minimum grid size on the predicted bubble motion for $(Bo, Ga) = (0.05, 90)$. Red and green lines correspond to cases with $\overline\Delta_{\min} = 1/68$ and $1/136$, respectively, while the blue line refers to the case where $\overline\Delta_{\min}$ is decreased from $1/68$ to $1/136$ only when $\overline\delta_{\min} \leq 0.15$. ($a$) and ($b$): bubble path in the wall-normal and wall-parallel planes, respectively; ($c$) and ($d$): evolution of the wall-normal ($V_x$) and vertical ($V_y$) velocities of the bubble centroid, respectively.}
\label{fig:pretest1-delta_min}
\end{figure}

Figure \ref{fig:pretest1-delta_min} compares the predicted bubble motion for $(Bo, Ga) = (0.05, 90)$ using three different settings for $\overline\Delta_{\min}$. For all three runs, $\zeta_u$ is fixed to $10^{-2}$. Irrespective of $\overline\Delta_{\min}$, the predicted lateral motion remains largely within the wall-normal plane, as the deviation of the bubble in the spanwise wall-parallel direction remains small, with $Z_b$ never exceeding $0.2$. The farthest wall-normal position where the bubble finally rests is about $6.75$ with $\overline\Delta_{\min}=1/68$, while it is about $5.5$, i.e., significantly closer to the wall, with $\overline\Delta_{\min}=1/136$. Close inspection of the predictions reveals that, as $\overline\Delta_{\min}$ decreases from $1/68$ to $1/136$, the bubble reaches a slightly larger separation during the final reversal stage (inset in panel ($a$)), resulting in a lower peak in the departure velocity as it escapes from the near-wall region (consider the situation at $T\approx12$ in panel ($c$)). The smaller final separation obtained with the finer grid follows. Conversely, the final rise speeds obtained with the different $\overline\Delta_{\min}$ show no discernible difference (panel ($d$)). These observations prove the importance of a sufficient spatial resolution within the wall boundary layer, especially during stages of intense bubble-wall interaction, to accurately predict the final wall-normal position of the bubble. This is further corroborated by the good agreement observed between predictions with $\overline\Delta_{\min}=1/136$ and those adopting an adaptive $\overline\Delta_{\min}$, where $\overline\Delta_{\min}=1/68$ is decreased to $1/136$ only during stages when $\overline\delta_{\min}(T) \leq 0.15$.

\begin{figure}
\centerline{\includegraphics[scale=0.65]{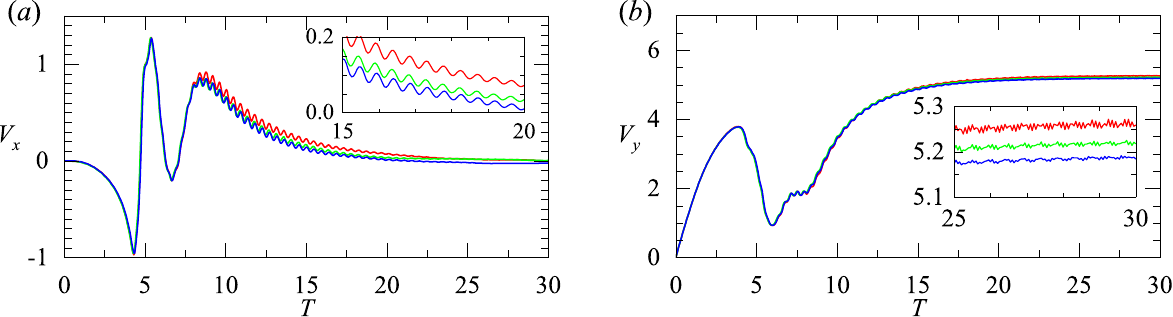}}
\vspace{-1mm}
\caption{Effects of $\zeta_u$ on the predicted wall-normal ($V_x$) and vertical ($V_y$) velocities of the bubble centroid for $(Bo, Ga) = (0.05, 90)$. Red, green and blue lines correspond to predictions obtained with $\zeta_u=10^{-2}$, $5\times10^{-3}$ and $2\times10^{-3}$, respectively. }
\label{fig:pretest1-zeta_u}
\vspace{-3mm}
\end{figure}

Effects of $\zeta_u$ are assessed based on the same test case. For this purpose, $\overline\Delta_{\min}$ is set to $1/68$ and $\zeta_u$ is decreased from $10^{-2}$ to $2\times10^{-3}$, which decreases the cell size within the far wake from $1/17$ to $1/34$. With this refinement, the total number of grid cells when the boundary layer at the bubble surface is fully developed increases from $4.3$ million to $21.8$ million. Figure \ref{fig:pretest1-zeta_u} compares the predictions obtained by varying $\zeta_u$ for the wall-normal and vertical bubble velocity components. It appears that the coarser resolution in the far wake leads to an underestimate of viscous effects. Specifically, the departing velocity (and to some minor extent the terminal rise speed) predicted with $\zeta_u=10^{-2}$ is slightly larger than that obtained using smaller values of $\zeta_u$. In contrast, predictions obtained with $\zeta_u=5\times10^{-3}$ and $\zeta_u=2\times10^{-3}$ exhibit only modest differences. Given this finding, and considering the significantly larger number of grid cells required with $\zeta_u=2\times10^{-3}$ ($21.8$ million compared with $8.4$ million), we set $\zeta_u=5\times10^{-3}$ in all runs discussed in the paper.

Following the above observations and taking advantage of our prior preliminary tests in Part 1, we find it reasonable to set the five numerical parameters as follows: $N_{CFL} = 0.5$, $T_\varepsilon = 10^{-4}$, $\zeta_f = 10^{-3}$, $\zeta_u = 5\times10^{-3}$, and $\overline\Delta_{\min} = 1/68$, automatically reduced to  $\overline\Delta_{\min} =1/136$ when $\overline\delta_{\min}(T) \leq 0.15$. 

\begin{figure}
\centerline{\includegraphics[scale=0.65]{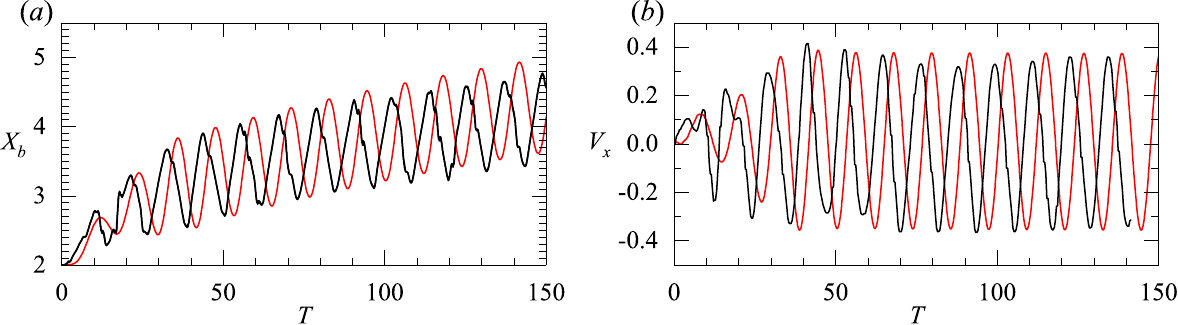}}
\vspace{-1mm}
\caption{Comparison between predictions (red line, $(Bo, Ga) = (1, 35)$) and experimental results from \cite{2024_Estepa-Cantero} (black line, $(Bo, Ga) = (0.97, 35)$) for the evolution of ($a$): the wall-normal position of the bubble centroid; and ($b$): the wall-normal velocity of the centroid.}
\label{fig:pretest2}
\end{figure}

To further check the numerical approach, we carry out two additional runs belonging to different $(Bo, Ga)$ ranges, for which reference data are available.
First, we select $(Bo, Ga) = (1, 35)$, corresponding to an air bubble with $R\approx 1.46\,\text{mm}$ rising in silicone oil DMS-T05 ($Mo=6.2\times10^{-7}$), as considered experimentally by \cite{2024_Estepa-Cantero}. Present predictions and experimental observations both conclude that the bubble departs from the wall and exhibits path oscillations in the wall-normal plane. Figure \ref{fig:pretest2} shows how the results for the evolution of the wall-normal bubble position, $X_b$, and velocity, $V_x$, compare. Owing to differences in the bubble shape and rise speed in the initial state, differences are observed between the predicted and the experimental evolutions in the early stages of the near-wall motion. The agreement is fairly good in later stages, although slight deviations subsist in the maximum and minimum wall-normal position beyond $T\approx100$. In the simulation, the Reynolds number based on the mean rise speed reached by the bubble when it moves far away from the wall ($X_b\geq 3$) is $\Rey\approx110.6$, closely matching the value determined experimentally in the absence of the wall, $\Rey=114.6$. Additionally, for $T\geq100$, the crest-to-crest amplitude of the transverse oscillations, the magnitude of the maximum wall-normal velocity, and the reduced frequency of the oscillations are $\overline{a}=1.37$, $V_x^{\max}=0.370$, and $St=0.108$ in the simulation, closely aligning with the experimental values $\overline{a}=1.36$, $V_x^{\max}=0.358$, and $St=0.104$.

\begin{figure}
\centerline{\includegraphics[scale=0.65]{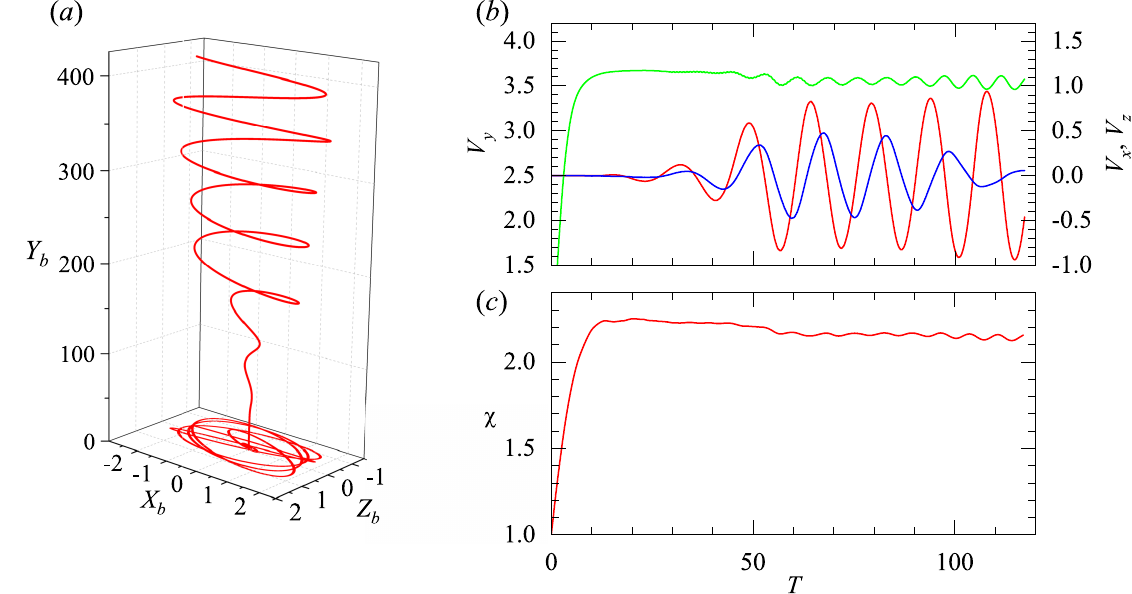}}
\caption{Predictions of the path characteristics of a single bubble with $(Bo, Ga) = (0.134, 99)$ rising from rest in an unbounded fluid domain. ($a$): front and bottom views of the path; ($b$): evolution of the vertical (green line), and horizontal (red and blue lines) velocity components of the bubble centroid; ($c$): evolution of the bubble aspect ratio. The path eventually converges to a planar zigzagging motion in the $(x,y)$ plane.}
\label{fig:pretest3}
\end{figure}

In the second case, we consider a bubble rising from rest in an unbounded domain for $(Bo, Ga) = (0.134, 99)$. This parameter set corresponds to an air bubble with $R=1~\text{mm}$ rising in pure water at $20\,^\circ\text{C}$ ($Mo=2.54\times10^{-11}$), a configuration already considered in experiments \citep{1995_Duineveld, 2014_Tagawa}. 
Figure \ref{fig:pretest3} illustrates the evolution of the bubble path, velocity and aspect ratio, all obtained from the present simulation. The bubble successively follows a straight path, then a flattened helical path, and finally tends to transition to a planar zigzagging path. This transition sequence differs from that reported by \cite{2014_Tagawa}, presumably due to the different initial conditions in the simulation and the experiment \citep{2015_Tchoufag}. Aside from this difference, the mean values for the rise speed $V_y$, the reduced frequency associated with the wall-normal velocity  $V_x$ (which is half that associated with $V_y$), and the bubble aspect ratio (averaged over the last five periods of oscillation of $V_y$) are $V_y=3.55,~ St=0.04,~\upchi=2.14$, all in good agreement with the values from previous investigations, namely $V_y=3.59, ~St=0.038, ~\upchi=2.14$.

\section{Influence of initial separation}
\label{sec:app_ini_sep}
The regime map gathering the various styles of path that are observed (figure \ref{fig:traj_sum}($a$)) was established based on simulations with an initial bubble-wall separation $X_0=2$. However, as the following discussion shows, this regime map remains unchanged when the initial separation is varied, provided that $X_0$ is not too large for wall effects to remain sizeable.

\begin{figure}
\centerline{\includegraphics[scale=0.65]{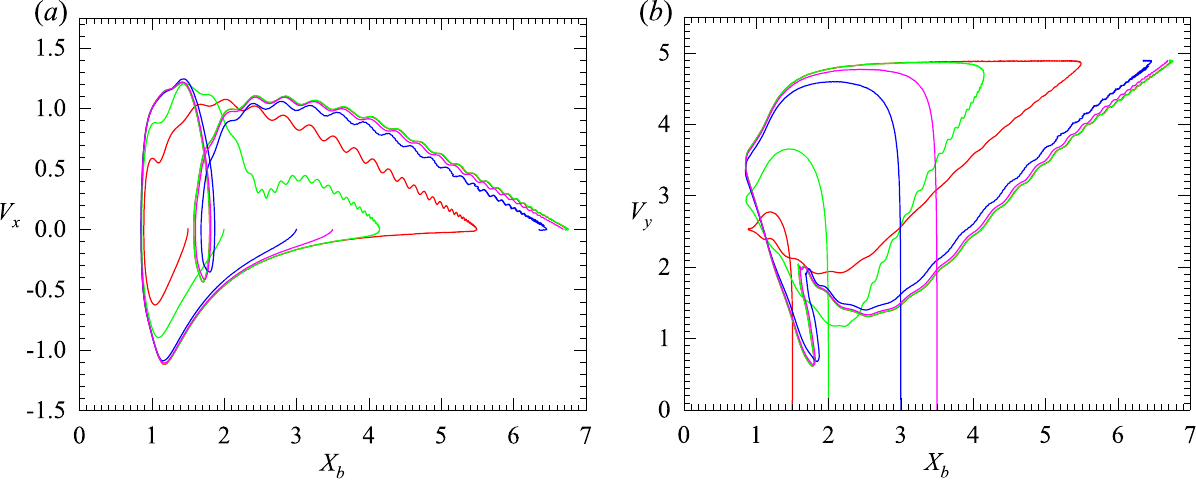}}
\vspace{-2mm}
\caption{Influence of the initial separation, $X_0$, on the evolution of the two components of the centroid velocity in the BTE regime for a bubble with $(Bo, Ga) = (0.05, 70)$. $(a)$: wall-normal velocity, $V_x$; $(b)$: vertical velocity, $V_y$. Red, green, blue and magenta lines correspond to initial separations $X_0 = 1.5, 2.0, 3.0$ and $3.5$, respectively. The red and green curves overlap in the second half of the evolution.}
\label{fig:escape-ini-sep}
\end{figure}

Below the neutral curve of path instability, four distinct scenarios take place. The influence of the initial separation on the first three of them (periodic near-wall bouncing, damped bouncing, migration away from the wall) was assessed in Part 1 (see appendix C therein). In all cases it was found that $X_0$ only affects the initial stage of the bubble motion, leaving it unchanged  in the fully developed state. Here, while $Ga$ is larger, the mechanisms responsible for these three scenarios remain unchanged and so are the conclusions reached in Part 1. 
Figure \ref{fig:escape-ini-sep} displays the influence of the initial separation in a case typical of the last of these four scenarios, the BTE regime. The corresponding $(Bo,Ga)$ set is that considered in figure \ref{fig:tre_motion}. All bubbles are found to eventually escape from the wall, since the final bubble-wall separation is large ($X_b(T\rightarrow\infty)>6$) in all cases. The only noticeable effect of $X_0$ is that the bubble needs to bounce twice against the wall before escaping when it is released close enough to wall, typically for $X_0\leq2.0$. This is because the maximum wall-normal distance the bubble attains after a bounce depends largely on its impact velocity, which itself depends on the magnitude of the attractive force resulting from the irrotational Bernoulli mechanism. As figure \ref{fig:escape-ini-sep}($b$) shows, the maximum rise speed achieved by the bubble before its first bounce decreases significantly with the initial separation when $X_0 \leq 2$, making the attractive force resulting from this mechanism (which goes like $V_y^2$) too weak to generate a large enough impact velocity during the first bounce in such cases.

\begin{figure}
\centerline{\includegraphics[scale=0.65]{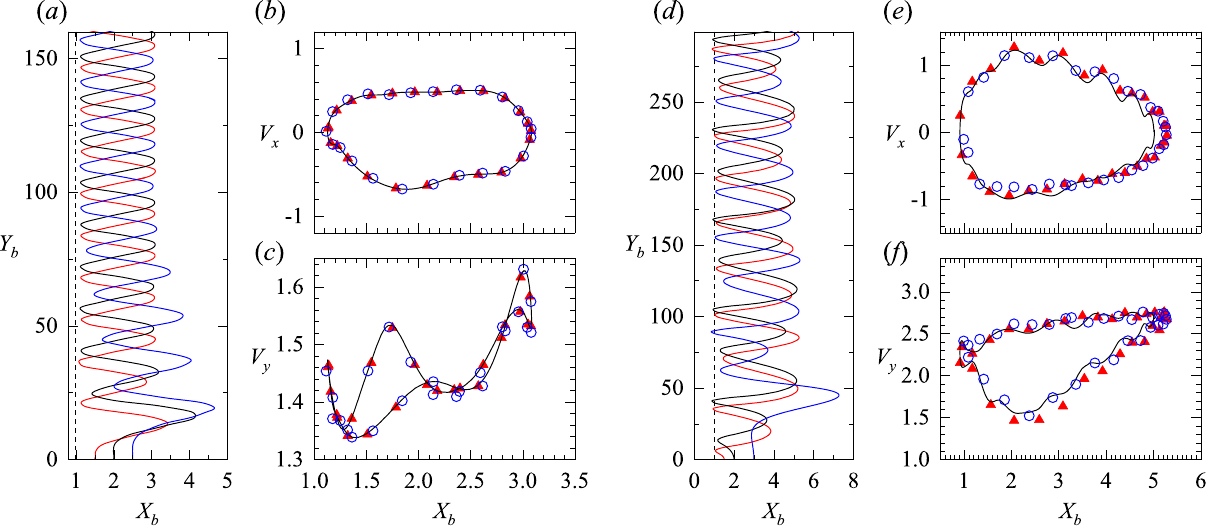}}
\vspace{-2mm}
\caption{Trajectories and velocities of bubbles released at different initial separations. $(a-c)$: $(Bo, Ga) = (1, 70)$; $(d-f)$: $(Bo, Ga) = (0.25, 90)$. Black line: $X_0=2$; red line and triangles: $X_0=1.5$; blue line and circles: $X_0=2.5$ in $(a-c)$ and $X_0=3$ in $(d-f)$. The time interval between adjacent points in $(b-c, e-f)$ is 0.5. Evolutions in $(b-c)$ correspond to the last zigzagging period, while in $(e-f)$ they refer to the last cycle of path oscillations when the bubble collides with the wall.  }
\label{fig:initial-seperation}
\end{figure}

For $(Bo,Ga)$ sets located above the neutral curve, bubbles follow either a WMA or a NWZ scenario (figure \ref{fig:traj_sum}($a$)). In the former case, the wall dictates the orientation of the plane in which the bubble oscillates in the early stages. Nevertheless, once the saturated state is reached, the characteristics of the motion are no longer affected by $X_0$. To reach this conclusion, we examined the two typical NWZ cases discussed in \S~\ref{sec:nwz}, comparing results obtained with three different initial separations. Figure \ref{fig:initial-seperation} shows the evolution of the bubble trajectory and velocity in these two cases. In the first of them (figure \ref{fig:initial-seperation}($a-c$)), direct collisions with the wall do not take place. As the corresponding panels show, the path characteristics always converge towards a developed state where the vertical displacement and amplitude of the lateral drift during a single period no longer vary over the zigzagging period. Comparing the evolutions obtained with different $X_0$ confirms that the path characteristics in this fully developed state are independent of $X_0$. Accordingly, the evolutions of the wall-normal and vertical velocities of the bubble centroid collapse on a single curve irrespective of $X_0$ (figure \ref{fig:initial-seperation}($b-c$)). The initial separation only influences the time required to reach the developed state. Specifically, at the largest separation ($X_0=2.5$), five periods of transverse oscillations take place before the developed state is reached. This is way slower than in a periodic near-wall bouncing scenario where the developed state is reached after the second period of bouncing irrespective of $X_0$. This indicates a relatively long memory of the system with respect to the initial separation for bubbles evolving in the NWZ regime. This memory effect becomes more pronounced when $Ga$ is large enough for bubble-wall collisions to take place (figure \ref{fig:initial-seperation}($d-f$)). With $X_0=2$, the bubble path reaches a fully developed state starting from the second bounce. In this developed stage, the amplitude of the lateral drift just after the collision is larger than that after the next or previous cycle of bouncing, where no collision takes place. In contrast, with $X_0=1.5$ and $3$, a much larger number of zigzagging periods is required to reach a similar state (figure \ref{fig:initial-seperation}$(d)$). The evolution of the two components of the bubble velocity in the last zigzagging cycle where collision takes place is reported in panels ($e-f$). While the evolutions corresponding to the three different $X_0$ are close, there are still sizeable differences, which highlights the long memory of the system with respect to $X_0$ at large $Ga$ and low-to-moderate $Bo$. 

The non-negligible influence of $X_0$ has been reported in previous experiments with air bubbles rising near a wall in pure water \citep{2015_Jeong, 2023_Cai}. These experiments were performed using bubbles with radii ranging from $1.14$ to $1.96\,$mm, which corresponds to $120\lesssim Ga\lesssim270$. In most cases, these bubbles were found to follow the NWZ scenario reported here. Nevertheless, given the larger $Ga$, the memory effect related to the initial separation may last for much longer times than in present simulations, such that very long vertical displacements are required for the path to reach a fully developed state. It is even plausible that in these experiments, the vertical distance crossed by the bubble before it reaches the measurement window ($\approx10R$ in \citet{2015_Jeong} and $\approx200R$ in \citet{2023_Cai}) is still not large enough for the corresponding memory effect to have completely faded away.

\bibliographystyle{jfm}
\bibliography{bubble-high-ga}

\begin{thebibliography}{45}
\expandafter\ifx\csname natexlab\endcsname\relax\def\natexlab#1{#1}\fi
\def\au#1{#1} \def\ed#1{#1} \def\yr#1{#1}\def\at#1{#1}\def\jt#1{\textit{#1}}
  \def\bt#1{#1}\def\bvol#1{\textbf{#1}} \def\vol#1{#1} \def\pg#1{#1}
  \def\publ#1{#1}\def\arxiv#1{#1}\def\org#1{#1}\def\st#1{\textit{#1}}

\bibitem[Auguste \& Magnaudet(2018)]{2018_Auguste}
{\sc \au{Auguste, F.} \& \au{Magnaudet, J.}} \yr{2018}  \at{Path oscillations
  and enhanced drag of light rising spheres}.  \jt{J. Fluid Mech.}  \bvol{841},
   \pg{228--266}.

\bibitem[Bonnefis {\em et~al.\/}(2023)Bonnefis, Fabre \&
  Magnaudet]{2023_Bonnefis}
{\sc \au{Bonnefis, P.}, \au{Fabre, D.} \& \au{Magnaudet, J.}} \yr{2023}
  \at{When, how, and why the path of an air bubble rising in pure water becomes
  unstable}.  \jt{Proc. Natl. Acad. Sci. U.S.A.}  \bvol{120},
  \pg{e2300897120}.

\bibitem[Bonnefis {\em et~al.\/}(2024)Bonnefis, Sierra-Ausin, Fabre \&
  Magnaudet]{2024_Bonnefis}
{\sc \au{Bonnefis, P.}, \au{Sierra-Ausin, J.}, \au{Fabre, D.} \& \au{Magnaudet,
  J.}} \yr{2024}  \at{Path instability of deformable bubbles rising in
  {N}ewtonian liquids: A linear study}.  \jt{J. Fluid Mech.}  \bvol{980},
  \pg{A19}.

\bibitem[Cai {\em et~al.\/}(2023)Cai, Ju, Chen \& Sun]{2023_Cai}
{\sc \au{Cai, R.}, \au{Ju, E.}, \au{Chen, W.} \& \au{Sun, J.}} \yr{2023}
  \at{Different modes of bubble migration near a vertical wall in pure water}.
  \jt{Korean J. Chem. Eng.}  \bvol{40},  \pg{67--78}.

\bibitem[Cano-Lozano {\em et~al.\/}(2016)Cano-Lozano, Mart\'inez-Baz\'an,
  Magnaudet \& Tchoufag]{2016_Cano-Lozano}
{\sc \au{Cano-Lozano, J.~C.}, \au{Mart\'inez-Baz\'an, C.}, \au{Magnaudet, J.}
  \& \au{Tchoufag, J.}} \yr{2016}  \at{Paths and wakes of deformable nearly
  spheroidal rising bubbles close to the transition to path instability}.
  \jt{Phys. Rev. Fluids}  \bvol{1},  \pg{053604}.

\bibitem[Duineveld(1994)]{1994_Duineveld}
{\sc \au{Duineveld, P.~C.}} \yr{1994}  \at{Bouncing and coalescence of two
  bubbles in water}. PhD thesis, Univ. Twente.

\bibitem[Duineveld(1995)]{1995_Duineveld}
{\sc \au{Duineveld, P.~C.}} \yr{1995}  \at{The rise velocity and shape of
  bubbles in pure water at high {Reynolds} number}.  \jt{J. Fluid Mech.}
  \bvol{292},  \pg{325--332}.

\bibitem[Ellingsen \& Risso(2001)]{2001_Ellingsen}
{\sc \au{Ellingsen, K.} \& \au{Risso, F.}} \yr{2001}  \at{On the rise of an
  ellipsoidal bubble in water: oscillatory paths and liquid-induced velocity}.
  \jt{J. Fluid Mech.}  \bvol{440},  \pg{235--268}.

\bibitem[Estepa-Cantero {\em et~al.\/}(2024)Estepa-Cantero, Mart\'inez-Baz\'an
  \& Bola\~nos Jim\'enez]{2024_Estepa-Cantero}
{\sc \au{Estepa-Cantero, C.}, \au{Mart\'inez-Baz\'an, C.} \& \au{Bola\~nos
  Jim\'enez, R.}} \yr{2024}  \at{Paths and wakes of deformable nearly
  spheroidal rising bubbles close to the transition to path instability}.
  \jt{Phys. Fluids}  \bvol{36},  \pg{013304}.

\bibitem[Horowitz \& Williamson(2010)]{2010_Horowitz}
{\sc \au{Horowitz, M.} \& \au{Williamson, C.~H.~K.}} \yr{2010}  \at{The effect
  of {Reynolds} number on the dynamics and wakes of freely rising and falling
  spheres}.  \jt{J. Fluid Mech.}  \bvol{651},  \pg{251--294}.

\bibitem[Jeong \& Park(2015)]{2015_Jeong}
{\sc \au{Jeong, H.} \& \au{Park, H.}} \yr{2015}  \at{Near-wall rising behaviour
  of a deformable bubble at high {Reynolds} number}.  \jt{J. Fluid Mech.}
  \bvol{771},  \pg{564--594}.

\bibitem[Jeong \& Hussain(1995)]{1995_Jeong}
{\sc \au{Jeong, J.} \& \au{Hussain, F.}} \yr{1995}  \at{On the identification
  of a vortex}.  \jt{J. Fluid Mech.}  \bvol{285},  \pg{69--94}.

\bibitem[Kok(1993)]{1993_Kok}
{\sc \au{Kok, J.~B.}} \yr{1993}  \at{Dynamics of a pair of gas bubbles moving
  through liquid. {P}art {I}. {T}heory.}  \jt{Eur. J. Mech. B Fluids}
  \bvol{12},  \pg{515--540}.

\bibitem[Lamb(1932)]{1932_Lamb}
{\sc \au{Lamb, H.}} \yr{1932} {\em Hydrodynamics\/}.  \publ{Cambridge
  University Press}.

\bibitem[Lee \& Park(2017)]{2017_Lee}
{\sc \au{Lee, J.} \& \au{Park, H.}} \yr{2017}  \at{Wake structures behind an
  oscillating bubble rising close to a vertical wall}.  \jt{Int. J. Multiphase
  Flow}  \bvol{91},  \pg{225--242}.

\bibitem[Magnaudet(2011)]{2011_Magnaudet}
{\sc \au{Magnaudet, J.}} \yr{2011}  \at{A reciprocal theorem for the prediction
  of loads on a body moving in an inhomogeneous flow at arbitrary {R}eynolds
  number}.  \jt{J. Fluid Mech.}  \bvol{689},  \pg{564--604}.

\bibitem[Magnaudet \& Mougin(2007)]{2007_Magnaudet}
{\sc \au{Magnaudet, J.} \& \au{Mougin, G.}} \yr{2007}  \at{Wake instability of
  a fixed spheroidal bubble}.  \jt{J. Fluid Mech.}  \bvol{572},  \pg{311--337}.

\bibitem[Meiron(1989)]{1989_Meiron}
{\sc \au{Meiron, D.~I.}} \yr{1989}  \at{On the stability of gas bubbles rising
  in an inviscid fluid}.  \jt{J. Fluid Mech.}  \bvol{198},  \pg{101--114}.

\bibitem[Miloh(1977)]{1977_Miloh}
{\sc \au{Miloh, T.}} \yr{1977}  \at{Hydrodynamics of deformable contiguous
  spherical shapes in an incompressible inviscid fluid}.  \jt{J. Eng. Math.}
  \bvol{11},  \pg{349--372}.

\bibitem[Moore(1963)]{1963_Moore}
{\sc \au{Moore, D.~W.}} \yr{1963}  \at{The boundary layer on a spherical gas
  bubble}.  \jt{J. Fluid Mech.}  \bvol{16},  \pg{161--176}.

\bibitem[Mougin \& Magnaudet(2002)]{2002b_Mougin}
{\sc \au{Mougin, G.} \& \au{Magnaudet, J.}} \yr{2002}  \at{Path instability of
  a rising bubble}.  \jt{Phys. Rev. Lett.}  \bvol{88},  \pg{014502}.

\bibitem[Mougin \& Magnaudet(2006)]{2006_Mougin}
{\sc \au{Mougin, G.} \& \au{Magnaudet, J.}} \yr{2006}  \at{Wake-induced forces
  and torques on a zigzagging/spiralling bubble}.  \jt{J. Fluid Mech.}
  \bvol{567},  \pg{185--194}.

\bibitem[Mundhra {\em et~al.\/}(2023)Mundhra, Lakkaraju, Das, Pakhomov \&
  Lobanov]{mundhra2023effect}
{\sc \au{Mundhra, R.}, \au{Lakkaraju, R.}, \au{Das, P.~K.}, \au{Pakhomov,
  M.~A.} \& \au{Lobanov, P.~D.}} \yr{2023}  \at{Effect of wall proximity and
  surface tension on a single bubble rising near a vertical wall}.  \jt{Water}
  \bvol{15},  \pg{1567}.

\bibitem[Popinet(2009)]{2009_Popinet}
{\sc \au{Popinet, S.}} \yr{2009}  \at{An accurate adaptive solver for
  surface-tension-driven interfacial flows}.  \jt{J. Comput. Phys.}
  \bvol{228},  \pg{5838--5866}.

\bibitem[Popinet(2015)]{2015_Popinet}
{\sc \au{Popinet, S.}} \yr{2015}  \at{A quadtree-adaptive multigrid solver for
  the {S}erre--{G}reen--{N}aghdi equations}.  \jt{J. Comput. Phys.}
  \bvol{302},  \pg{336--358}.

\bibitem[Popinet(2017)]{2017_Popinet}
{\sc \au{Popinet, S.}} \yr{2017}  \at{Bubble rising in a large tank.} Available
  at: \url{http://basilisk.fr/src/examples/bubble.c}.

\bibitem[Rastello {\em et~al.\/}(2009)Rastello, Mari{\'e}, Grosjean \&
  Lance]{rastello2009drag}
{\sc \au{Rastello, M.}, \au{Mari{\'e}, J.~L.}, \au{Grosjean, N.} \& \au{Lance,
  M.}} \yr{2009}  \at{Drag and lift forces on interface-contaminated bubbles
  spinning in a rotating flow}.  \jt{J. Fluid Mech.}  \bvol{624},
  \pg{159--178}.

\bibitem[Shi(2024)]{2024_Shi_a}
{\sc \au{Shi, P.}} \yr{2024}  \at{Reversal of the transverse force on a
  spherical bubble rising close to a vertical wall at moderate-to-high
  {Reynolds} numbers}.  \jt{Phys. Rev. Fluids}  \bvol{9},  \pg{023601}.

\bibitem[Shi {\em et~al.\/}(2020)Shi, Rzehak, Lucas \& Magnaudet]{2020_Shi}
{\sc \au{Shi, P.}, \au{Rzehak, R.}, \au{Lucas, D.} \& \au{Magnaudet, J.}}
  \yr{2020}  \at{Hydrodynamic forces on a clean spherical bubble translating in
  a wall-bounded linear shear flow}.  \jt{Phys. Rev. Fluids}  \bvol{5},
  \pg{073601}.

\bibitem[Shi {\em et~al.\/}(2024)Shi, Zhang \& Magnaudet]{2024_Shi_b}
{\sc \au{Shi, P.}, \au{Zhang, J.} \& \au{Magnaudet, J.}} \yr{2024}  \at{Lateral
  migration and bouncing of a deformable bubble rising near a vertical wall.
  {P}art 1. {M}oderately inertial regimes}.  \jt{J. Fluid Mech.}  \bvol{998},
  \pg{A8}.

\bibitem[Sugioka \& Tsukada(2015)]{2015_Sugioka}
{\sc \au{Sugioka, K.~I.} \& \au{Tsukada, T.}} \yr{2015}  \at{Direct numerical
  simulations of drag and lift forces acting on a spherical bubble near a plane
  wall}.  \jt{Int. J. Multiphase Flow}  \bvol{71},  \pg{32--37}.

\bibitem[Tagawa {\em et~al.\/}(2014)Tagawa, Takagi \& Matsumoto]{2014_Tagawa}
{\sc \au{Tagawa, Y.}, \au{Takagi, S.} \& \au{Matsumoto, Y.}} \yr{2014}
  \at{Surfactant effect on path instability of a rising bubble}.  \jt{J. Fluid
  Mech.}  \bvol{738},  \pg{124--142}.

\bibitem[Takemura \& Magnaudet(2003)]{2003_Takemura}
{\sc \au{Takemura, F.} \& \au{Magnaudet, J.}} \yr{2003}  \at{The transverse
  force on clean and contaminated bubbles rising near a vertical wall at
  moderate {Reynolds} number}.  \jt{J. Fluid Mech.}  \bvol{495},
  \pg{235--253}.

\bibitem[Takemura {\em et~al.\/}(2002)Takemura, Takagi, Magnaudet \&
  Matsumoto]{2002_Takemura}
{\sc \au{Takemura, F.}, \au{Takagi, S.}, \au{Magnaudet, J.} \& \au{Matsumoto,
  Y.}} \yr{2002}  \at{Drag and lift forces on a bubble rising near a vertical
  wall in a viscous liquid}.  \jt{J. Fluid Mech.}  \bvol{461},  \pg{277--300}.

\bibitem[Tchoufag {\em et~al.\/}(2015)Tchoufag, Fabre \&
  Magnaudet]{2015_Tchoufag}
{\sc \au{Tchoufag, J.}, \au{Fabre, D.} \& \au{Magnaudet, J.}} \yr{2015}
  \at{Weakly nonlinear model with exact coefficients for the fluttering and
  spiraling motion of buoyancy-driven bodies}.  \jt{Phys. Rev. Lett.}
  \bvol{115},  \pg{114501}.

\bibitem[Veldhuis(2007)]{veldhuis2007leonardo}
{\sc \au{Veldhuis, C.}} \yr{2007}  \at{Leonardo's paradox: Path and shape
  instabilities of particles and bubbles}. PhD thesis, Univ. Twente.

\bibitem[Veldhuis {\em et~al.\/}(2008)Veldhuis, Biesheuvel \& van
  {Wijngaarden}]{2008_Veldhuis}
{\sc \au{Veldhuis, C.}, \au{Biesheuvel, A.} \& \au{van {Wijngaarden}, L.}}
  \yr{2008}  \at{Shape oscillations on bubbles rising in clean and in tap
  water}.  \jt{Phys. Fluids}  \bvol{20},  \pg{040705}.

\bibitem[de~{Vries}(2001)]{2001_Vries}
{\sc \au{de~{Vries}, A. W.~G.}} \yr{2001}  \at{Path and wake of a rising
  bubble}. PhD thesis, Univ. Twente.

\bibitem[de~{Vries} {\em et~al.\/}(2002)de~{Vries}, Biesheuvel \& van
  {Wijngaarden}]{2002_Vries}
{\sc \au{de~{Vries}, A. W.~G.}, \au{Biesheuvel, A.} \& \au{van {Wijngaarden},
  L.}} \yr{2002}  \at{Notes on the path and wake of a gas bubble rising in pure
  water}.  \jt{Int. J. Multiphase Flow}  \bvol{28},  \pg{1823--1835}.

\bibitem[van {Wijngaarden}(1976)]{1976_Wijngaarden}
{\sc \au{van {Wijngaarden}, L.}} \yr{1976}  \at{Hydrodynamic interaction
  between gas bubbles in liquid}.  \jt{J. Fluid Mech.}  \bvol{77},
  \pg{27--44}.

\bibitem[Yan {\em et~al.\/}(2022)Yan, Zhang, Liao, Zhang, Zhou \&
  Liu]{2022_Yan}
{\sc \au{Yan, H.}, \au{Zhang, H.}, \au{Liao, Y.}, \au{Zhang, H.}, \au{Zhou, P.}
  \& \au{Liu, L.}} \yr{2022}  \at{A single bubble rising in the vicinity of a
  vertical wall: A numerical study based on volume of fluid method}.  \jt{Ocean
  Eng.}  \bvol{263},  \pg{112379}.

\bibitem[Zenit \& Magnaudet(2008)]{2008_Zenit}
{\sc \au{Zenit, R.} \& \au{Magnaudet, J.}} \yr{2008}  \at{Path instability of
  rising spheroidal air bubbles: {A} shape-controlled process}.  \jt{Phys.
  Fluids}  \bvol{20},  \pg{061702}.

\bibitem[Zenit \& Magnaudet(2009)]{2009_Zenit_a}
{\sc \au{Zenit, R.} \& \au{Magnaudet, J.}} \yr{2009}  \at{Measurements of the
  streamwise vorticity in the wake of an oscillating bubble}.  \jt{Int. J.
  Multiphase Flow}  \bvol{35},  \pg{195--203}.

\bibitem[Zhang {\em et~al.\/}(2021)Zhang, Ni \& Magnaudet]{2021_Zhang}
{\sc \au{Zhang, J.}, \au{Ni, M.} \& \au{Magnaudet, J.}} \yr{2021}
  \at{Three-dimensional dynamics of a pair of deformable bubbles rising
  initially in line. {P}art 1. {M}oderately inertial regimes}.  \jt{J. Fluid
  Mech.}  \bvol{920},  \pg{A16}.

\bibitem[Zhang {\em et~al.\/}(2020)Zhang, Dabiri, Chen \& You]{2020_Zhang}
{\sc \au{Zhang, Y.}, \au{Dabiri, S.}, \au{Chen, K.} \& \au{You, Y.}} \yr{2020}
  \at{An initially spherical bubble rising near a vertical wall}.  \jt{Int. J.
  Heat Fluid Flow}  \bvol{85},  \pg{108649}.

\end{thebibliography}
\end{document}